\documentclass[sigconf, nonacm]{acmart}
\pdfoutput=1
\usepackage{listings}
\usepackage{graphicx}

\usepackage{float}
\newfloat{algorithm}{tb}{lop}

\usepackage{stfloats}
\usepackage{caption}
\usepackage{subcaption}
\usepackage{xspace}
\usepackage{booktabs}
\usepackage{dcolumn}
\usepackage{tabularx}
\usepackage{mathtools}
\usepackage[binary-units]{siunitx}
\usepackage{pifont}
\usepackage{environ}
\usepackage[htt]{hyphenat}
\usepackage[colorinlistoftodos,prependcaption,textsize=tiny]{todonotes}

\usepackage{enumitem}
\setlist{nosep}

\usepackage{tikz}
\usetikzlibrary{calc}
\usetikzlibrary{fit}
\usetikzlibrary{positioning}
\usetikzlibrary{shapes.symbols}
\usepackage{pgfplots}
\usetikzlibrary{shapes, arrows, positioning, fit, calc, decorations.markings,
decorations.pathmorphing, shadows, backgrounds, positioning, patterns}
\tikzstyle{vecArrow} = [thick, decoration={markings,mark=at position
   1 with {\arrow[semithick]{open triangle 60}}},
   double distance=1.4pt, shorten >= 5.5pt,
   preaction = {decorate},
   postaction = {draw,line width=1.4pt, white,shorten >= 4.5pt}]

\tikzstyle{snakeArrow} = [->,thick, decorate,decoration={snake,amplitude=.4mm,
    segment length=2mm,post length=1mm}]
\tikzstyle{innerWhite} = [semithick, white,line width=1.4pt, shorten >= 4.5pt]
\tikzstyle{decision} = [diamond, text centered, text width=2cm,
                        draw=black, fill=green!30]
\tikzstyle{block} = [rectangle, draw, fill=blue!20,
    text width=4em, text centered, rounded corners]
\tikzstyle{smallblock} = [rectangle, draw, fill=blue!20, text centered, rounded corners]
\tikzstyle{sblock} = [rectangle, draw, fill=blue!20,
    text centered, rounded corners, minimum height=4em]
\tikzstyle{initjoinblock} = [rectangle, draw, fill=gray,
    text centered, minimum height=2em, minimum width=2em]
\tikzstyle{joinblock} = [rectangle, draw, fill=gray!20,
    text centered, minimum height=4em, minimum width=2em]
\tikzstyle{jitqblock} = [rectangle, draw, fill=blue!20,
    text centered, rounded corners, align=center, rectangle
    split, rectangle split parts=2]
\tikzstyle{bigblock} = [rectangle, draw, fill=blue!20,
    text width=\linewidth, text centered, rounded corners, minimum height=4em]
\tikzstyle{arrow} = [thick,->,>=stealth]
\tikzstyle{lightarrow} = [draw=gray,->,>=stealth]

\tikzset{halo/.style={preaction={draw,white,line width=4pt,
            -
        },
        preaction={draw,
            white,
            ultra thick,
            shorten >=-2.5\pgflinewidth
        }
    }
}
\tikzset{%
  cascaded/.style = {%
    general shadow = {%
      shadow scale = 1,
      shadow xshift = -2ex,
      shadow yshift = 2ex,
      draw,
      fill=white,
      thick},
    general shadow = {%
      shadow scale = 1,
      shadow xshift = -1ex,
      shadow yshift = 1ex,
      draw,
      fill=white,
      thick,
      },
    draw,
    fill=white,
    thick,
    minimum width = 1.5cm,
    minimum height = 2cm}}

\makeatletter
\newcommand{\currentplotnumber}{\the\pgfplots@numplots}
\newcommand\numberofbarplots{0}
\makeatother

\pgfkeys{/pgfplots/highlight missing data/.style={
    filter point/.code = {
      \pgfkeys{/pgf/fpu}
      \pgfmathparse{\pgfkeysvalueof{/data point/y}}
      \pgfmathfloattofixed{\pgfmathresult}
      \pgfkeys{/pgf/fpu=false}
      \pgfmathparse{\pgfmathresult==#1}
      \ifnum\pgfmathresult=1
        \pgfkeys{/pgfplots/x coord inv trafo=\pgfkeysvalueof{/data point/x}}
        \let\xposition=\pgfmathresult
        \pgfmathparse{max(\numplotsofactualtype,\numberofbarplots)}
        \global\let\numberofbarplots=\pgfmathresult
        \edef\mystar{
          \noexpand\node at (axis cs:\xposition,0.5) [xshift=-0.5*(2pt+\noexpand\pgfplotbarwidth)*\noexpand\numberofbarplots+1*(\currentplotnumber*(2pt+\noexpand\pgfplotbarwidth)),anchor=base] {*}; 
        }
        \mystar
        \pgfkeys{/data point/y=nan}
      \else
        \pgfmathparse{\pgfkeysvalueof{/data point/y}}
      \fi
      }
  },/pgfplots/highlight missing data/.default=-1 
}

\DeclareSIUnit{\nothing}{\relax}
\DeclareSIUnit{\x}{\times}
\DeclareSIUnit{\ct}{\text{\textcent}}
\DeclareSIUnit\invocations{\text{invocations}}

\makeatletter
\DeclareRobustCommand{\varname}[1]{\begingroup\newmcodes@\mathit{#1}\endgroup}

\definecolor{ETHc}{RGB}{18,105,176}     
\newenvironment{revision}
  {\color{black}%
   }
 {}

\newcommand\vldbdoi{10.14778/3484224.3484229}
\newcommand\vldbpages{3308 - 3321}
\newcommand\vldbvolume{14}
\newcommand\vldbissue{13}
\newcommand\vldbyear{2021}

\newcommand\vldbtitle{\shorttitle}
\newcommand\vldbavailabilityurl{}
\newcommand\vldbpagestyle{empty}

\begin{document}
\title[Modularis: Modular Relational Analytics over Heterogeneous Distributed Platforms]{Modularis: Modular Relational Analytics \\ over Heterogeneous Distributed Platforms}

\author{Dimitrios Koutsoukos}
\affiliation{%
  \state{ETH Zurich}
  \country{Switzerland}
}
\email{dkoutsou@inf.ethz.ch}

\author{Ingo M{\"u}ller}
\affiliation{%
  \state{ETH Zurich}
  \country{Switzerland}
}
\email{ingo.mueller@inf.ethz.ch}

\author{Renato Marroqu{\'i}n$^*$}
\affiliation{%
  \state{Oracle Inc.}
  \country{Zurich, Switzerland}
}
\email{renato.marroquin@oracle.com}

\author{Ana Klimovic}
\affiliation{%
  \state{ETH Zurich}
  \country{Switzerland}
}
\email{aklimovic@ethz.ch}

\author{Gustavo Alonso}
\affiliation{%
  \state{ETH Zurich}
  \country{Switzerland}
}
\email{alonso@inf.ethz.ch}
%

\renewcommand{\shortauthors}{Dimitrios Koutsoukos, Ingo M{\"u}ller, Renato
    Marroqu{\'i}n, Ana Klimovic, Gustavo Alonso}
\begin{abstract}
The enormous quantity of data produced every day together with advances in
data analytics has led to a proliferation of data management and analysis
systems. Typically, these systems are built around highly specialized
monolithic operators optimized for the underlying hardware.
While effective in the short term, such an approach
makes the operators cumbersome to port and adapt, 
which is increasingly required due to the speed at which algorithms and hardware
evolve. To address this limitation, we present \emph{Modularis}, %
an execution layer for data analytics based on \emph{sub-operators}, i.e., composable building blocks resembling traditional database operators
but at a finer granularity.
To demonstrate the feasibility and advantages of our approach, we use Modularis
to build a distributed query processing system supporting relational queries
running on an RDMA cluster, a serverless cloud platform,
\begin{revision}and a smart storage engine.\end{revision}
Modularis requires minimal code changes to execute queries across these three diverse hardware platforms,
showing that the sub-operator approach reduces the amount and complexity of the
code to maintain. In fact, changes in the platform affect only those sub-operators
that depend on the underlying hardware (in our use cases, mainly the
sub-operators related to network communication). We show the
end-to-end performance of Modularis by comparing it with a framework for
SQL processing (Presto), a commercial cluster database (SingleStore),
as well as Query-as-a-Service systems (Athena, BigQuery).
Modularis outperforms all these systems, proving that the design and
architectural advantages of a modular design can be achieved without degrading
performance. We also compare Modularis with a hand-optimized implementation of
a join for RDMA clusters. We show that Modularis has the advantage of being
easily extensible to a wider range of join variants and \textit{group by}
queries, all of which are not supported in the hand-tuned join.
\end{abstract}

\maketitle

\pagestyle{\vldbpagestyle}
\begingroup\small\noindent\raggedright\textbf{PVLDB Reference Format:}\\
\shortauthors. \vldbtitle. PVLDB, \vldbvolume(\vldbissue): \vldbpages, \vldbyear.\\
\href{https://doi.org/\vldbdoi}{doi:\vldbdoi}
\endgroup
\begingroup
\renewcommand\thefootnote{}\footnote{\noindent
*The work of this author was done while employed at ETH Zurich.\\
This work is licensed under the Creative Commons BY-NC-ND 4.0 International License. Visit \url{https://creativecommons.org/licenses/by-nc-nd/4.0/} to view a copy of this license. For any use beyond those covered by this license, obtain permission by emailing \href{mailto:info@vldb.org}{info@vldb.org}. Copyright is held by the owner/author(s). Publication rights licensed to the VLDB Endowment. \\
\raggedright Proceedings of the VLDB Endowment, Vol. \vldbvolume, No. \vldbissue\ %
ISSN 2150-8097. \\
\href{https://doi.org/\vldbdoi}{doi:\vldbdoi} \\
}\addtocounter{footnote}{-1}\endgroup

\ifdefempty{\vldbavailabilityurl}{}{
\vspace{.3cm}
\begingroup\small\noindent\raggedright\textbf{PVLDB Artifact Availability:}\\
The source code, data, and/or other artifacts have been made available at \url{\vldbavailabilityurl}.
\endgroup
}

\section{Introduction}
The growing popularity of machine learning applications and the increasing
amount of data that analytics applications must process
have had a substantial influence on the way systems are designed and
optimized. There is a constant stream of specialized accelerators
(TPUs, GPUs, FPGAs, smart NICs, smart storage, near memory processing)
and platforms (large appliances, InfiniBand clusters, serverless, cloud instances,
data centers) that forces a continuous redesign of data processing engines---often leading to new engines---simply to exploit the capabilities of new hardware~\cite{binnig2016end}.

Often, to gain performance, developers design monolithic operators that are highly tailored
to the underlying hardware \cite{BalkesenTAO13,Barthels2015,Barthels2017,perron2020starling,muller2020lambada}. However, as the algorithms and platforms evolve quickly, it becomes very difficult to reuse these operators
in newer versions of the system. Examples abound: For instance, a join optimized for
multi-core machines~\cite{BalkesenTAO13} requires fundamental changes
to run on Remote Direct Memory Access (RDMA)~\cite{Barthels2015, Barthels2017} due to
the different communication schemes between NUMA nodes and
the network. As another example,
although FPGAs are not competitive with multi-core machines for full joins,
they can significantly accelerate the partitioning phase~\cite{kaan17}.
Supporting distributed query processing on serverless computing has received a lot of recent attention and requires a specialized exchange operator to allow communication through
storage~\cite{muller2020lambada,perron2020starling}.
With the current approach of highly engineered, monolithic operators, it
is difficult to exploit the potential of new
architectures and platforms without major redesigns.

We argue that to cope with the fast changes in the hardware and
platform landscape, query processing needs to become more
modular and composable at a finer granularity than conventional relational
operators. Having many versions of highly optimized, monolithic
operators is not a design approach compatible with the high degree
of specialization we observe. In almost all cases where hardware or
platform advances offer new opportunities, the potential advantages affect only
part of an operator (e.g., only one of partitioning, build, or phase of a join)
and often require to change other significant parts (e.g., intermediate data
placement in an exchange operator, partitioning strategies, etc.).
It is very rare that the whole operator can become
faster or that all operators benefit. This effect is
notable in accelerators
(FPGA, GPU)~\cite{SidlerIOA17,HeSIA18,fang2019accelerating,dennl2012fly,dursun2019morsel,
Teubner2011,Polychroniou2015,Govindaraju2006,He2008,He2009}, specialized
processor components (AVX, SGX),~\cite{BalkesenTAO13,AgrawalIRVGVBGR17,
kim2009sort,Muller2015,Cieslewicz2007,Schuhknecht2015}
evolving networks in distributed systems (Infiniband, RDMA, smart NICs,
etc.)~\cite{Barthels2015,zamanian2019rethinking,
li2016accelerating,liu2017design,ziegler2019designing,Blanas2020,
rodiger2015high}, and even platforms (Infiniband
clusters~\cite{Barthels2015,li2016accelerating,liu2017design},
cloud/serverless computing~\cite{muller2020lambada,fouladi2019laptop,jonas2017occupy,
klimovic2018pocket,ao2018sprocket,kim2018serverless,pu2019shuffling,
sampe2018serverless,shankar2018numpywren,carreira2019cirrus,
klimovic2018understanding,perron2020starling}).
Yet, the current design approaches often
require a complete redesign because they are
based on careful tailoring to the underlying platform and hardware.

In this paper, we focus on the modular design of query processing. We show how to design a modular distributed
query processing engine with performance comparable to its monolithic
counterparts. The topic of modularity in database operators has been
visited many times in the past~\cite{dittrich2000component, irmert2008new}
and recently~\cite{dittrich2019case, kohn2021building}. However, to our knowledge,
we are the first to implement modularity at the hardware platform level, while at the same
time formulating concrete design principles. We argue that modularity at this level is a
necessity rather than a nice-to-have feature.
To this end, we have built \emph{Modularis}---an execution engine
aiming to maximize performance without specializing neither the engine nor the
bulk of the operators to the target platform. Modularis is based on a
collection of composable sub-operators that are both as small and simple
as possible, as well as reusable, while retaining the ability to execute entire
SQL queries. Modularis' sub-operators share the same goal of composability
of operators present in traditional database engines: sub-operators can be
freely and easily combined, have a well-defined interface, and adhere to a common
execution model. Our sub-operators are similar to the microcode used in
processor design to implement more complex instructions. We use them to build up complex plans like TPC-H
queries. \begin{revision} This allows Modularis to run seamlessly on three different platforms:
an InfiniBand RDMA cluster, a serverless cloud
service, and a smart storage engine\end{revision}, by simply replacing in the
query plan only those operators actually affected by the change in the platform
(e.g., the exchange operator), leaving everything else unaffected.

We have evaluated Modularis' performance and behavior extensively.
First, to explore the end-to-end performance of Modularis, we compare it to
mature systems using the TPC-H benchmark. When compared to Presto, a system that
is general enough to utilize various storage layers and distributed set-ups, Modularis is
an order of magnitude faster. When compared to SingleStore (previously called MemSQL), a system that specializes in
in-memory analytics using SQL, Modularis is faster for the majority of the queries. The speed-up
compared to both of these systems comes from the ability to
optimize at the sub-operator level and the usage of RDMA for fast data transfer.
Next, we show how Modularis adapts to heterogeneous
environments by discussing the minimal changes necessary to go from running on
an RDMA cluster to a serverless platform using AWS Lambda
\begin{revision} or a smart storage engine (S3Select) \end{revision} ---%
a very significant architectural change in the underlying platform.
For TPC-H queries, Modularis-on-serverless is competitive against
commercial Query-as-a-Service systems (Athena, BigQuery). This last experiment
shows that modularity does not need to result in end-to-end performance losses
while it enables the execution of workloads on two fundamentally different
platforms with minimal development effort, which mainly involves the design of
new exchange and executor operators. That way, we get the best of both worlds:
maximum performance across platforms, without redesigning the whole system from
scratch.

Second, we quantify the potential performance overhead of the modular design
by comparing a join composed from several Modularis' sub-operators with a
hand-tuned join. For the latter, we use the best monolithic implementations
available for RDMA clusters~\cite{Barthels2015,Barthels2017}.
Modularis is always within \SI{30}{\percent} of the performance
of the specialized implementation (and often closer).
Nevertheless, our system uses \SI{3.8}{\times} fewer lines of code
than the monolithic operator and all its sub-operators are not specific
to the join but can be reused in other query plans.

Third, we demonstrate the advantages of sub-operators over monolithic approaches
when it comes to extending existing operators. We show how to use the same
sub-operators that we used for the join
for optimizations for sequences of joins
and a distributed \texttt{GROUP BY} (with one additional sub-operator).
In contrast, extending existing manually handcrafted
joins~\cite{Barthels2015,Barthels2017,liu2017design}
to support, e.g., inner, outer, semi, and anti joins plus grouping
for partitioned, sorted, and general inputs would be very
difficult (and has not been attempted, to our knowledge).
In Modularis, once we had the sub-operators for the initial distributed
radix hash join, we needed only a small effort to develop the rest of the
operators. These results put into perspective the potential performance loss
when comparing Modularis, which can run code over different platforms,
with handcrafted algorithms tailored to run on a single target system.


\section{Related work}\label{sec:related_work}

\begin{description}[style=multiline,wide,nosep]
\item[Database operators design.]
Operator modularity is a crucial design decision as it significantly affects
performance.
Determining the right granularity of operators
has been a reoccurring topic of research:
from the bracket model~\cite{Dewitt1990gamma}
for parallelization in the early days of databases
to objected-oriented modular designs~\cite{dittrich2000component},
record-oriented components adapted at runtime~\cite{irmert2008new},
morsel-driven parallelism~\cite{Leis14},
and ``deep'' query optimization~\cite{dittrich2019case} more recently.
In Modularis, we use the Volcano
model~\cite{Graefe1990} as the basis for the interfaces between operators, but
we also support collections instead of just flat records.
Modularis shares a similar vision to that of Dittrich and
Nix~\cite{dittrich2019case, mutable}, Bandle and
Giceva~\cite{bandle2021database}, and Kohn et al.~\cite{kohn2021building}
in having operators defined at a finer granularity.
Dittrich and Nix argue that
this enables a deeper level of query
optimization and easy implementation of research ideas without sacrificing
performance. Bandle and Giceva analyze how sub-operators can be used for general
data analytics. Both these works sketch a vision for modular operator design,
and they focus on a few variations of aggregation using
different indexes (sorting/hashing) or on how to build more complex algorithms
(e.g hash joins, k-means) out of sub-operators. In contrast, Modularis is a full
system comprising a variety of
operators and runs full TPC-H queries using plans tailored to three very different platforms.
Kohn et al.~\cite{kohn2021building} develop a framework that uses
modularity to speed up query execution when the query contains multiple
aggregates. Modularis follows a similar spirit
but builds a more generic execution layer that achieves a similar goal
for query execution in general. In contrast to these approaches,
we tackle the problem at the
platform level and formulate design principles for these operators. We also
do not focus on the optimization aspect from the view of query rewriting.
Our goal is to reduce the implementation effort and keep up with the fast pace
with which the hardware evolves, without having to rewrite systems completely.
Finally, efforts seeking to optimize
 sets of operators are orthogonal to Modularis
as they could be applied to the query plan generated by
Modularis as additional optimization passes.
For instance, Leeka et al.~\cite{leeka19} fuse similar
operators into a single \textit{super-operator} by
using a streaming interface.

\item[Emerging technologies in data processing.]
Two of the most recent emerging technologies in distributed data processing are
RDMA and serverless computing.
On the one hand, RDMA has been used to implement highly distributed
relational operators~\cite{liu2017design,Barthels2015, Barthels2017} and
several projects have explored
the network bottleneck and the use of RDMA for query processing and database
design in a variety of contexts~\cite{Barthels2019,rodiger2015high,
salama2017rethinking,binnig2016end,li2016accelerating}.
On the other hand, serverless computing has been extensively studied
over the last few years for a
variety of applications~\cite{fouladi2019laptop,jonas2017occupy,
klimovic2018pocket,klimovic2018understanding,ao2018sprocket,kim2018serverless,
pu2019shuffling,sampe2018serverless,shankar2018numpywren,carreira2019cirrus,
perron2020starling}.
In both cases, research focused on solving platform-specific challenges
and was often carried out in built-from-scratch prototypes
that were limited in functionality.
In Modularis, by leveraging the granularity of the sub-operators, we develop
specialized platform-specific operators that leverage the latest technologies
of both platforms but keep most of
the other operators and the rest of the system hardware-agnostic.
This shows that we can achieve competitive performance
without having to redesign the entire system.

\item[Query compilation for data processing.]
Modularis is also related in part to systems that perform Just-in-Time (JiT) query
compilation as pioneered in HyPer \cite{Neumann2011hyper}. Similar techniques were later used and extended in
several other systems: Tupleware~\cite{Crotty2015tupleware}, which targets
machine learning workloads; Weld~\cite{Palkar2018weld}, which incorporates a
large class of data analytics algorithms by translating frontend languages into
an intermediate representation (IR) that is later Just-in-Time compiled;
LegoBase~\cite{legobase2014}, which builds a database using a high-level
language; and Flare~\cite{flare2018,flare_lantern2019}, which combines data
processing tasks with machine learning. Modularis uses JiT compilation
techniques similar to those in such systems to eliminate the potential performance
overhead of a modular design. The systems above either have an IR
tailored towards single-machine execution and rely on systems such as Spark~\cite{zaharia2010spark} for
their distributed setup (e.g. Weld) or, if they have a distributed setting
(e.g. Tupleware), they use a very generic model that is very hard to be adapted in
case of platform changes (e.g. from VMs to serverless functions). Modularis on
the other hand, targets directly distributed analytics and focuses on
tailoring execution to the underlying platforms by only using minimal changes to the plans.
\end{description}

\section{Modular operators}\label{sec:modular_operators}

\subsection{Modularis architecture}\label{subsec:system_arch}

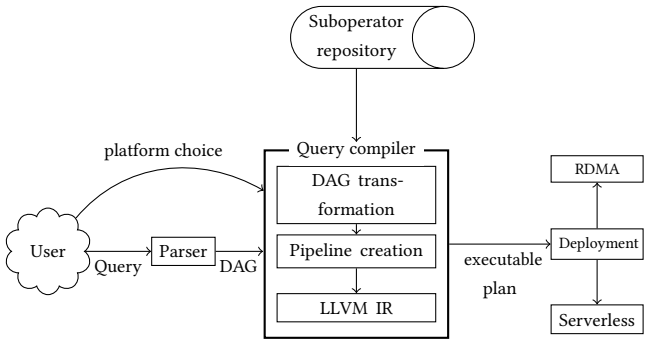
\begin{figure}[t]
   \centering
   \begin{tikzpicture}[node distance=1.75cm, auto]
       \node [cloud, draw=black, text width=1.5em, align=center, xshift=-2em] (user)
           {\footnotesize{User}};
       \node [rectangle, right of=user, draw=black, node distance=1.8cm,
              text width=2em, align=center] (parser)
              {\footnotesize{Parser}};
       \node [rectangle, right of=parser, draw=black, node distance=2.3cm,
              text width=6em, align=center] (pipeline)
              {\footnotesize{Pipeline creation}};
       \node [rectangle, above of=pipeline, draw=black, node distance=0.75cm,
              text width=6em, align=center] (transformation)
              {\footnotesize{DAG transformation}};
       \node [rectangle, below of=pipeline, draw=black, node distance=0.75cm,
              text width=6em, align=center] (llvm)
              {\footnotesize{LLVM IR}};
       \node [draw, thick, inner xsep=0.5em, inner ysep=0.65em, fit=
            (pipeline)(llvm)(transformation)] (compiler) {};
       \node[fill=white] at (compiler.north) {\footnotesize{Query compiler}};
       \node [cylinder, above of=compiler, draw=black, node distance=2.75cm,
              text width=4em, align=center] (repository)
              {\footnotesize{Suboperator repository}};
       \node [rectangle, right of=compiler, draw=black, node distance=3.2cm,
              text width=3.2em, align=center] (deployment)
              {\scriptsize{Deployment}};
       \node [rectangle, above of=deployment, draw=black, node distance=1cm,
              text width=3.2em, align=center] (rdma)
              {\scriptsize{RDMA}};
       \node [rectangle, below of=deployment, draw=black, node distance=1cm,
              text width=3.2em, align=center] (serverless)
              {\footnotesize{Serverless}};
          \draw[->] (user) -- (parser) node [midway, below] {\footnotesize{Query}};
        \draw[->, shorten >=0.18cm] (parser) -- (pipeline) node [midway, below,
            xshift=-0.1cm] {\footnotesize{DAG}};
        \draw[->] (compiler) -- (deployment) node [midway,
            below, text width=3em, align=center]{\footnotesize{executable plan}};
        \draw [->] (user) to [out=50,in=150]  node [midway,
            above] {\footnotesize{platform choice}} (compiler);
        \draw[->] (transformation) -- (pipeline);
        \draw[->] (pipeline) -- (llvm);
        \draw[->, shorten >=0.1cm] (repository) -- (compiler);
        \draw[->] (deployment) -- (rdma);
        \draw[->] (deployment) -- (serverless);
   \end{tikzpicture}
   \caption{System architecture}\label{fig:system}
   \vspace{-2em}
\end{figure}

In this section, we describe the architecture of Modularis and how it
executes a query. We show the system architecture in
Figure~\ref{fig:system}.
\begin{revision}The user writes queries in a UDF-based library interface written in
Python (similar to PySpark)\end{revision}. When the user submits the query,
she specifies the target execution platform using a flag
(e.g. ,\texttt{-{}-rdma}, \texttt{-{}-lambda}).
The query is then parsed and translated into an IR,
representing a DAG of operators. \begin{revision}
The DAG is serialized and passed to Modularis' backend, written in C++,
which applies a series of both plan-specific and platform-specific
transformations.\end{revision}

The plan-specific transformations involve projection and selection push-downs,
operations typical in DBMSs to reduce data movement and I/O.
Next, Modularis cuts the DAG into tree-shaped sub-plans,
each of which represents a pipeline with a materialization point at its end.
Then, we transform the query into its distributed equivalent. This step
involves wrapping the initial plan into a distributed executor and adding
exchange operators in the plan inputs. Depending on the target
platform, we specialize the generic operators with hardware-specific ones.
In the case of serverless, we use the exchange operator of
Lambada~\cite{muller2020lambada} and an executor that spawns the workers in a
tree-plan fashion. In the case of smart storage, we use a specialized operator
to get data from S3Select. In the case of RDMA, we use an MPI executor and an
RDMA-based exchange operator based on the one of Barthels
et. al~\cite{Barthels2017} (with modifications to avoid unnecessary data movement).
We give
concrete examples of the final plan in Sections
\ref{subsec:tpch-queries} and \ref{subsec:smart_storage}. Finally, the
plan is lowered into LLVM IR and Just-in-Time compiled
to native machine code. To translate UDFs into LLVM IR, Modularis uses
Numba~\cite{numba} and inlines the generated LLVM code into the remainder of the
plan to eliminate any function calls or interpretation in
inner loops. The query is then executed on the target platform,
based on the specialized operator that it has been wrapped around.
When the results are produced, they are returned back to the user.

\vspace{-2mm}
\subsection{Design principles}\label{sec:design-principles}
Modularis builds on the observation
that the commonly used operators,
e.g., for high-speed networks~\cite{Barthels2015}
or multi-core CPUs~\cite{BalkesenTAO13},
are built around the same \emph{conceptual} building blocks:
when the authors describe their algorithms,
they use some visual or textual representation of
``reading data'', ``partition by key'', ``for each partition'', \mbox{etc.
To} readers, these terms imply the same operation is being done
during different phases of query execution.
However, there is no code reuse---%
the implementations differ in intricacies related to how and where data is
stored, how it is passed from one phase to the next,
how they depend on the state of some enclosing scope, etc.

The goal of Modularis is to identify pieces of code that reoccur in operators in slight variations,
factor out their common logic, and package them behind a well-defined interface
such that they can be reused and recomposed.
In other words, our goal is to derive \emph{actual} building blocks
from the conceptual ones.
To derive sub-operators systematically,
we follow these design principles:
\begin{enumerate}[wide,nosep]
  \item \emph{Each sub-operator consists of or is a part of
              at most one inner loop.}
    If a high-level algorithm consists of several phases,
    each of these phases is expressed by at least one sub-operator.
    Phases often reoccur across and within monolithic operators,
    sometimes in slight variations. For example, join operators typically use
    partitioning to improve cache locality. In
    Section~\ref{sec:complex-plans:group-by}, we show that by factoring out the
    partitioning logic into a dedicated sub-operator, we can reuse it to
    improve cache locality in grouping operators as well.

\item \emph{Use dedicated sub-operators for each physical (in-memory) data
        materialization format.}
    This decouples the processing of data from where and how it is stored.
    Consequently, other sub-operators become independent
    of the physical formats of their inputs and outputs
    and are more generic.
    One high-level example is to have different scan sub-operators
    for reading base tables or intermediate materializations in RDMA buffers.
    That way, a single partitioning sub-operator implementation
    can consume inputs of two different scan operators (or any other operator)
    instead of having two specialized partitioning operators
    (see Section~\ref{sec:complex-plans:partitioned-hash-join}).

  \item \emph{Express high-level control flow as (nested) operators. }
    This allows connecting plan fragments of sub-operators
    that express the heavy-lifting data processing
    through the same operator interface.
    In monolithic operators, such orchestration logic is usually implemented
    as imperative code specific to that operator
    and, thus, it makes it necessary
    to reimplement the data path as imperative code as well.
    One high-level example is to express the in-memory join
    of two partition pairs occurring in a classical partitioned hash join
    as a nested query plan
    that is executed for each pair of matching partitions. That
    allows the use of partition-unaware sub-operators in the inner plan.
    We introduce the \texttt{NestedMap} sub-operator for this purpose below
    (Section~\ref{sec:complex-plans:partitioned-hash-join}).
\end{enumerate}

\begin{revision}
\textbf{Generality of design principles. } The above somewhat different
implementations do not only apply to the context of
CPUs, but also to other hardware architectures (e.g. FPGAs, GPUs). Therefore,
the goal of the design principles is still applicable, as we can use them
to derive the basic building blocks that the underlying architectures use.
By following  our design principles, we do not expect to design operators that work solely in
different architectures without any modifications. We rather want
to avoid reimplementing the same conceptual building blocks with small
modifications. Our goal is to end up with a
very similar set of operators for different
architectures. For example, an operator that reads data in
columnar format is relevant to both CPUs and FPGAs. By having a similar set of
building blocks for different architectures, we can offload parts of a
query to different hardware configurations seamlessly, just by swapping out
the hardware-specific part of the plan. We give a concrete example of how this
can be done in the context of smart storage in Section~\ref{subsec:smart_storage}.
\end{revision}

\subsection{The sub-operator interface}
\label{subsec:interface}
We base the interface of sub-operators
on one of the most known models in the database community,
the Volcano model~\cite{Graefe1990}.
The model is based on iterators that pass records along the data path
of a tree of \texttt{Next()} function calls.
Like in a traditional execution engine,
the iterator interface allows us to combine operators
in almost arbitrary ways,
limited only by the schema or types that operators may require.

The main distinctive feature of the sub-operator interface
compared to traditional Volcano-style operators
is the type system of the records (or ``tuples'') passed between them.
While records in relations (in the First Normal Form)
consist of atomic fields,
we need a more expressive type system
for a generic physical execution layer.
For example, to split the materialize and scan operators
of a given physical data format into two distinct sub-operators,
these operators need to pass ``records'' containing the materialization
from one to the other.
Similarly, if we want to express operators that work on individual records
as well as those that work on batches (or morsels \cite{Leis14})
in the same interface,
we need to be able to represent the concept of a ``batch''.
We thus extend the concept of tuples with that of ``collections'',
which is the generalization
of any physical data format of tuples of a particular type.
This allows expressing physical execution properties into the query plan
rather than hard-coding them for the entire execution engine.

More formally, sub-operators are iterators over $\varname{tuples}$
and the tuples are of a statically known type
from the following recursive type structure:
\vspace{-2mm}
\begin{align*}
  \varname{tuple} &:=
    \langle\varname{item},\,\ldots,\,\varname{item}\rangle \\
  \varname{item} &:= \left\{\,
        \varname{atom} \,\middle|\,
        \varname{collection}~\text{of}~\varname{tuples}
    \,\right\},
\end{align*}
\vspace{-0.5mm}
where a $\varname{tuple}$ is a mapping
from a domain of (static) field identifiers to item types,
an $\varname{item}$ is a (statically known) atomic or collection-based type,
an $\varname{atom}$ is a particular domain of undividable values,
and a $\varname{collection}$ is the generalization
of any physical data format one might want to use in the execution layer.
We denote tuple types by
$\langle\varname{fieldName0}: \varname{ItemType0},\,\ldots,\,
\varname{fieldNameK}: \varname{ItemTypeK}\rangle$
and collection types by
$\varname{CollectionType}\langle\varname{TupleType}\rangle$.
Most operators are \emph{generic}
in the sense that they require their upstream operator(s)
to produce tuples of a type with a particular structure
but accept any type of that structure.
Their output type usually depends on the type(s) of their upstream(s).
For example, the scan operator for a C-array of C-structs
(which we call $\varname{RowVector}$)
requires from their upstreams to produce tuples of type
$\varname{RowVector}\langle\varname{TupleType}\rangle$
and returns tuples of $\varname{TupleType}$,
where $\varname{TupleType}$ is allowed to be any tuple type.
Similarly, operators consuming or producing batches
can do that by consuming or producing tuples with $\varname{RowVector}$ fields.

We also extend the Volcano-style execution model to DAGs, whereas operators in the original Volcano-style execution
model could only have one consumer (i.e., plans had to be trees).
Before execution, we cut a DAG of operators into pipelines,
where pipelines start with either the original plan inputs
or the result of any operator with several consumers.
In each pipeline, that result will be read only once,
so the sub-plan of the pipeline is a tree
and can thus be executed with the iterator model.
Pipelines materialize their results,
such that multiple downstream pipelines can read them.
For simplicity, we present the plans as DAGs in the remainder of the paper
and omit the pipelines and materialization points.

\subsection{Initial set of sub-operators}\label{sec:sub_operators}
In this section, we introduce the initial set of sub-operators that we use in
Modularis. \begin{revision} We choose the sub-operators
such that they are expressive enough
to support all the major relational operations, e.g.\
selections, projections, aggregations, and joins. At the same time, we
follow our design principles. While the design principle of making operators
simple aims indeed at increasing code re-usage and reducing implementation effort,
this does not necessarily mean that there is no redundancy. In fact, we do add new operators
if they provide a sufficiently large and broad performance benefit.
\end{revision}Such an
example is the different join
implementations (inner, semi, anti). They could all be based on the same hash-build and
a separate probe operator. However, having a special operator for each of them provides better performance.
Similarly, having a second version for each of them with flipped build and probe sides
increases performance further. \begin{revision}Finally, while the
initial set of sub-operators is enough for relational analytics, we
expect that, as we expand Modularis to more types of analytics (e.g. linear
algebra, ML), we need to expand our sub-operator set. \end{revision}

We present our initial list of sub-operators in Table~\ref{tab:operators}.
The sub-operators fall into six categories:
orchestration operators, data processing operators, MPI-specific operators (to support RDMA clusters),
Lambda-specific operators (to support serverless), \begin{revision}
Smart storage-specific operators\end{revision},
and generic materialize and scan operators.
Orchestration operators enable the execution of nested computations.
Data processing operators express the computations
carried out on the data inside the inner loops.
MPI- and Lambda-specific operators are the ones
that are aware of the distributed nature of query execution.
Smart storage-specific operators use pushdown computations to smart storage.
Finally, materialize and scan operators
read and write $\varname{tuples}$ from and to nested $\varname{collections}$.
We give an overview of our operators in Table~\ref{tab:operators}.
We believe that the semantics of most operators are clear given their names.
However, the two orchestration operators merit an explanation,
as they differ substantially from other systems.

\begin{table}[b]
    \begin{tabular}{|l|p{4cm}|}
    \hline
    \textbf{Category} & \textbf{Operators} \\
    \hline
    Orchestration operators & Parameter Lookup, NestedMap \\
    \hline
    Data processing operators & (Parametrized) Map, Projection, Cartesian
    Product, Filter, Reduce (By Key), GroupBy, Zip, Local Histogram,
    Build and Probe, Partition, Semi-join, Sort, Top-K  \\
    \hline
    MPI-specific operators & MPI Executor, MPI Histogram, MPI Exchange \\
    \hline
    Lambda-specific operators & Lambda Executor, Lambda Exchange \\
    \hline
    Smart storage-specific operators & S3Select Scan \\
    \hline
    Materialize and scan operators & Local Partitioning (AVX-based), Partition,
    Row Scan, Column Scan, Parquet Scan, Materialize Row Vector, Arrow table to collection \\
    \hline
    \end{tabular}
    \caption{Initial set of sub-operators}
    \label{tab:operators}
\vspace{-3mm}
\end{table}
The
\texttt{ParameterLookup} operator encapsulates plan inputs
in the operator interface,
such that other operators can consume them. This operator is the only operator
aware of plan inputs.
The \texttt{NestedMap} operator
executes a nested plan independently on each input tuple,
which typically contains a nested collection.
Each invocation of the nested plan produces an output tuple
that may contain nested collections as well.
This allows us to process nested collections
using the same building blocks regardless of the nesting level.
This operator consumes tuples of any type,
and the \texttt{ParameterLookup} operator(s) in the nested plan
return a tuple of that type.

\vspace{-2mm}
\section{From operators to complex query plans}\label{sec:complex_qplans}
\subsection{High-performance distributed join}\label{subsec:join-description}
We illustrate how Modularis' sub-operators can express
optimized monolithic operators
with a case study of a state-of-the-art distributed join algorithm
proposed by Barthels et al. \cite{Barthels2017}.
\subsubsection{State-of-the-art distributed join}
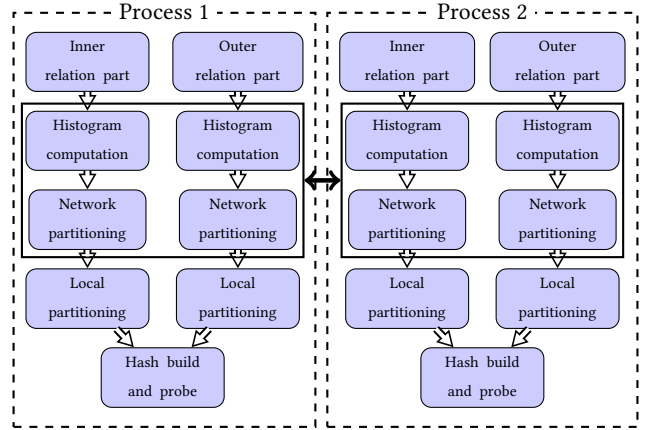
\begin{figure}[t]
  \centering
  \begin{tikzpicture}[node distance=1.05cm, auto]
  \node [block, xshift=-2cm, text width=4.5em] (inner)
    {\scriptsize{Inner relation part}};
  \node [block, text width=4.8em] (outer) {\scriptsize Outer relation part};
  \node [block, below of=inner, text width=4.5em] (hist1)
    {\scriptsize Histogram computation};
  \node [block, below of=outer, text width=4.5em] (hist2)
    {\scriptsize Histogram computation};
  \node [block, below of=hist1, text width=4.25em] (net1)
    {\scriptsize Network partitioning};
  \node [block, below of=hist2, text width=4.25em] (net2)
    {\scriptsize Network partitioning};
  \node [block, below of=net1, text width=4.5em] (local1)
    {\scriptsize Local partitioning};
  \node [block, below of=net2, text width=4.5em] (local2)
    {\scriptsize Local partitioning};
  \node [block, below of=local1, xshift=1cm, text width=4.5em] (hash)
    {\scriptsize Hash build and probe};
  \node [draw, thick, dashed, inner xsep=0.5em, inner ysep=0.8em, fit=
  (inner)(outer)(hist1)(hist2)(net1)(net2)(local1)(local2)(hash)] (proc1) {};
  \node[fill=white] at (proc1.north) {Process 1};

  \node [block, text width=5em, right of=inner, node distance=4.25cm] (inner2)
    {\scriptsize Inner relation part};
  \node [block, text width=5em, right of=inner2, node distance=2cm] (outer2)
    {\scriptsize Outer relation part};
   \node [block, below of=inner2, text width=4.5em] (hist3)
    {\scriptsize Histogram computation};
  \node [block, below of=outer2, text width=4.5em] (hist4)
    {\scriptsize Histogram computation};
  \node [block, below of=hist3, text width=4.25em] (net3)
    {\scriptsize Network partitioning};
  \node [block, below of=hist4, text width=4.25em] (net4)
    {\scriptsize Network partitioning};
  \node [block, below of=net3, text width=4.5em] (local3)
    {\scriptsize Local partitioning};
  \node [block, below of=net4, text width=4.5em] (local4)
    {\scriptsize Local partitioning};
  \node [block, below of=local3, xshift=1cm, text width=4.5em] (hash2)
    {\scriptsize Hash build and probe};
  \node [draw, thick, dashed, inner xsep=0.5em, inner ysep=0.8em,
         fit=(inner2)(outer2)(hist3)(hist4)(net3)(net4)(local3)(local4)(hash2)] (proc2) {};
  \node[fill=white] at (proc2.north) {Process 2};
  \node [draw, thick, inner xsep=0.15em, inner ysep=0.3em,
         fit=(net1)(net2)(hist1)(hist2)] (network) {};
  \node [draw, thick, inner xsep=0.15em, inner ysep=0.3em,
         fit=(net3)(net4)(hist3)(hist4)] (network2) {};

  \draw [vecArrow] (outer) -- (hist2);
  \draw [vecArrow] (inner) -- (hist1);
  \draw [vecArrow] (hist1) -- (net1);
  \draw [vecArrow] (hist2) -- (net2);
  \draw [vecArrow] (net1) -- (local1);
  \draw [vecArrow] (net2) -- (local2);
  \draw [vecArrow] (local1) -- (hash);
  \draw [vecArrow] (local2) -- (hash);
  \draw [vecArrow] (outer2) -- (hist4);
  \draw [vecArrow] (inner2) -- (hist3);
  \draw [vecArrow] (hist3) -- (net3);
  \draw [vecArrow] (hist4) -- (net4);
  \draw [vecArrow] (net3) -- (local3);
  \draw [vecArrow] (net4) -- (local4);
  \draw [vecArrow] (local3) -- (hash2);
  \draw [vecArrow] (local4) -- (hash2);
  \draw [line width=0.6mm, <->] (network)  -- (network2);
\end{tikzpicture}
  \caption{RDMA-aware hash join algorithm
    for two processes proposed by \cite{Barthels2017}}\label{fig:original-join-dag}
\vspace{-2em}
\end{figure}

We start with a very brief summary of the algorithm as it was originally
proposed (Figure~\ref{fig:original-join-dag}). The algorithm
consists of three phases:
(1) histogram computation,
(2) multi-pass partitioning including network transfer,
and (3) hash table build and probe.
The two phases where communication happens amongst processes,
namely the histogram calculation and the network partitioning phase,
are depicted using black boxes around them.
The original algorithm is optimized for a workload involving two relations
where both relations consist of 16-byte tuples
(8 bytes for the key and another 8 for the payload).
For more details, we refer the readers to the original paper~\cite{Barthels2017}.

\subsubsection{Query plan in Modularis}\label{sec:complex-plans:partitioned-hash-join}

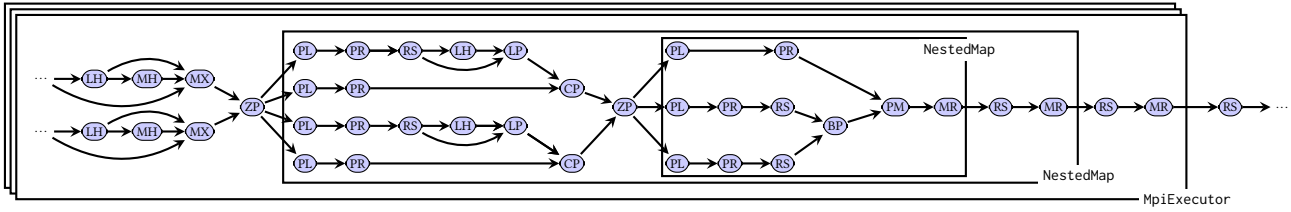
\begin{figure*}
    \centering
    \tiny
    \begin{tikzpicture}[node distance=0.7cm, auto]
    \node [rectangle, thick] (dots1) {\dots};
    \node [rectangle, thick, below of=dots1] (dots2) {\dots};
    \node [smallblock, right of=dots1] (lh1) {LH};
    \node [smallblock, right of=lh1] (gh1) {MH};
    \node [smallblock, right of=gh1] (nw1) {MX};
    \node [smallblock, right of=dots2] (lh2) {LH};
    \node [smallblock, right of=lh2] (gh2) {MH};
    \node [smallblock, right of=gh2] (nw2) {MX};
    \node [smallblock, right of=nw1, yshift=-0.375cm] (zip1) {ZP};
    \node [smallblock, right of=zip1, yshift=0.75cm] (pl3) {PL};
    \node [smallblock, right of=zip1, yshift=0.25cm] (pl4) {PL};
    \node [smallblock, right of=zip1, yshift=-0.25cm] (pl5) {PL};
    \node [smallblock, right of=zip1, yshift=-0.75cm] (pl6) {PL};
    \node [smallblock, right of=pl3] (pr1) {PR};
    \node [smallblock, right of=pl4] (pr2) {PR};
    \node [smallblock, right of=pl5] (pr3) {PR};
    \node [smallblock, right of=pl6] (pr4) {PR};
    \node [smallblock, right of=pr1] (rs1) {RS};
    \node [smallblock, right of=pr3] (rs2) {RS};
    \node [smallblock, right of=rs1] (lh3) {LH};
    \node [smallblock, right of=lh3] (lp1) {LP};
    \node [smallblock, below of=lp1, node distance=0.5cm, xshift=0.75cm] (cp1) {CP};
    \node [smallblock, right of=rs2] (lh4) {LH};
    \node [smallblock, right of=lh4] (lp2) {LP};
    \node [smallblock, below of=lp2, node distance=0.5cm, xshift=0.75cm] (cp2) {CP};
    \node [smallblock, right of=cp1, yshift=-0.25cm] (zip2) {ZP};
    \node [smallblock, right of=zip2, yshift=0.75cm] (pl7) {PL};
    \node [smallblock, right of=zip2] (pl8) {PL};
    \node [smallblock, right of=zip2, yshift=-0.75cm] (pl9) {PL};
    \node [smallblock, right of=pl7, xshift=0.75cm] (pr5) {PR};
    \node [smallblock, right of=pl8] (pr6) {PR};
    \node [smallblock, right of=pl9] (pr7) {PR};
    \node [smallblock, right of=pr6] (rs3) {RS};
    \node [smallblock, right of=pr7] (rs4) {RS};
    \node [smallblock, right of=rs3, yshift=-0.25cm] (hp) {BP};
    \node [smallblock, right of=rs3, node distance=1.5cm] (pm) {PM};
    \node [smallblock, right of=pm] (mrv1) {MR};
    \node [smallblock, right of=mrv1] (rs5) {RS};
    \node [smallblock, right of=rs5] (mrv2) {MR};
    \node [smallblock, right of=mrv2] (rs6) {RS};
    \node [smallblock, right of=rs6] (dots3) {MR};
    \begin{scope}[on background layer]
    \node [cascaded, inner xsep=1em, inner ysep=2em, fit=
    (dots1)(dots2)(nw1)(nw2)(zip1)(pl3)(pl4)(pl5)(pl6)(pr1)(pr2)(pr3)(pr4)(rs1)(rs2)(lp1)(cp1)(lp2)(cp2)(zip2)(pl7)(pl8)(pl9)(pr5)(pr6)(pr7)(rs3)(rs4)(hp)(pm)(mrv1)(mrv2)(rs5)(rs6)(dots3)] (inner1) {};
    \node[fill=white] at (inner1.south east) {\scriptsize \texttt{MpiExecutor}};
    \end{scope}
    \node [draw, thick, inner xsep=0.75em, inner ysep=0.75em, fit=
        (pl3)(pl4)(pl5)(pl6)(pr1)(pr2)(pr3)(pr4)(rs1)(rs2)(lh3)(lp1)(cp1)(lp2)(cp2)(zip2)(pl7)(pl8)(pl9)(pr5)(pr6)(pr7)(rs3)(rs4)(hp)(pm)(mrv1)(rs5)(mrv2)] (inner2) {};
    \node [draw, thick, inner xsep=0.25em, inner ysep=0.25em, fit=
    (pl7)(pl8)(pl9)(pr5)(pr6)(pr7)(rs3)(rs4)(hp)(pm)(mrv1)] (inner3) {};
    \node[fill=white, yshift=0.1cm] at (inner2.south east) {\scriptsize \texttt{NestedMap}};
    \node[fill=white, yshift=-0.17cm, xshift=-0.1cm] at (inner3.north east) {\scriptsize \texttt{NestedMap}};
    \node [smallblock, node distance=0.95cm, right of=dots3] (frs) {RS};
    \node [rectangle, thick, right of=frs] (dots4) {\dots};

    \draw [arrow] (dots1) -- (lh1);
    \draw [arrow] (lh1) -- (gh1);
    \draw [arrow] (dots1) to[bend right] (nw1);
    \draw [arrow] (lh1) to[bend left] (nw1);
    \draw [arrow] (gh1) -- (nw1);
    \draw [arrow] (dots2) -- (lh2);
    \draw [arrow] (lh2) -- (gh2);
    \draw [arrow] (dots2) to[bend right] (nw2);
    \draw [arrow] (lh2) to[bend left] (nw2);
    \draw [arrow] (gh2) -- (nw2);
    \draw [arrow] (nw1) -- (zip1);
    \draw [arrow] (nw2) -- (zip1);
    \draw [arrow] (zip1) -- (pl3);
    \draw [arrow] (zip1) -- (pl4);
    \draw [arrow] (zip1) -- (pl5);
    \draw [arrow] (zip1) -- (pl6);
    \draw [arrow] (pl3) -- (pr1);
    \draw [arrow] (pl4) -- (pr2);
    \draw [arrow] (pl5) -- (pr3);
    \draw [arrow] (pl6) -- (pr4);
    \draw [arrow] (pr1) -- (rs1);
    \draw [arrow] (pr1) -- (rs1);
    \draw [arrow] (pr3) -- (rs2);
    \draw [arrow] (rs1) to[bend right] (lp1);
    \draw [arrow] (rs1) -- (lh3);
    \draw [arrow] (lh3) -- (lp1);
    \draw [arrow] (rs2) to[bend right] (lp2);
    \draw [arrow] (lh4) -- (lp2);
    \draw [arrow] (rs2) -- (lh4);
    \draw [arrow] (lh4) -- (lp2);
    \draw [arrow] (pr2) -- (cp1);
    \draw [arrow] (lp1) -- (cp1);
    \draw [arrow] (pr4) -- (cp2);
    \draw [arrow] (lp2) -- (cp2);
    \draw [arrow] (lp2) -- (cp2);
    \draw [arrow] (cp1) -- (zip2);
    \draw [arrow] (cp2) -- (zip2);
    \draw [arrow] (zip2) -- (pl7);
    \draw [arrow] (zip2) -- (pl8);
    \draw [arrow] (zip2) -- (pl9);
    \draw [arrow] (pl7) -- (pr5);
    \draw [arrow] (pl8) -- (pr6);
    \draw [arrow] (pl9) -- (pr7);
    \draw [arrow] (pr6) -- (rs3);
    \draw [arrow] (pr7) -- (rs4);
    \draw [arrow] (rs3) -- (hp);
    \draw [arrow] (rs4) -- (hp);
    \draw [arrow] (pr5) -- (pm);
    \draw [arrow] (hp) -- (pm);
    \draw [arrow] (pm) -- (mrv1);
    \draw [arrow] (mrv1) -- (rs5);
    \draw [arrow] (rs5) -- (mrv2);
    \draw [arrow] (mrv2) -- (rs6);
    \draw [arrow] (rs6) -- (dots3);
    \draw [arrow] (dots3) -- (frs);
    \draw [arrow] (frs) -- (dots4);
\end{tikzpicture}
    \vspace{-2mm}
    \caption{Plan that runs the distributed hash join with modular operators across many nodes}\label{fig:modular_join}
    \vspace{-2mm}
\end{figure*}

We now show how the same join algorithm can be expressed using sub-operators in
our RDMA backend. The resulting query plan is shown in Figure~\ref{fig:modular_join}.
To keep the graphical representation concise,
we abbreviate the operator names as shown in Table~\ref{tab:legend}.
Furthermore, we omit materialization points
and, instead, express the plan as a DAG as discussed previously.
Finally, most of the operators in the Figure
are part of a nested plan inside a \texttt{MpiExecutor} operator,
which is executed concurrently
by all \begin{revision} MPI processes (which we call \textit{ranks} in the
rest of the section)\end{revision} in the cluster in a data-parallel way, illustrated with a stacked frame.

\begin{table}[b]
\centering
\vspace{-4mm}
\begin{tabular}{@{}c l r@{}} \toprule
    \textbf{Abbreviation} & \textbf{Operator name}  & \textbf{SLOC}\\ \midrule
    PL & Parameter lookup & 28\\
    NM & Nested map & 49\\
    \midrule
    PR & Projection & 27 \\
    BP & Hash build and probe & 103 \\
    LH & Local histogram & 77 \\
    ZP & Zip & 44 \\
    CP & Cartesian product & 54 \\
    PM & Parametrized map & 51 \\
    RK & Reduce by key & 75 \\
    \midrule
    RS & Row Scan & 59 \\
    LP & Local partitioning & 143\\
    MR & Materialize row vector & 56 \\
    \midrule
    ME & MPI Executor & 140 \\
    MX & MPI Exchange & 269\\
    MH & MPI Histogram & 52\\\bottomrule
\end{tabular}
\caption{Source line of code per operator}
\label{tab:legend}
\vspace{-2mm}
\end{table}

The join starts by computing the histogram of each of the two inputs
using the \texttt{LocalHistogram} operator.
The inputs can be produced by any operator producing tuples with key fields,
e.g., a scan operator reading from a base table stored in main memory.
On each of the two sides, a \texttt{MpiHistogram} computes the global
histogram from the local ones until the \texttt{MpiExchange} consumes
the local histograms, the global histogram, and the original input. With this
information, each rank allocates \begin{revision} a contiguous memory area in
the main memory of the host (called \textit{RMA window})\end{revision}
that will hold all tuples it will receive in this phase
and computes the offsets of its exclusive regions
inside the windows of the target ranks.
Then, each rank reads the input again,
computes the target partition for each input tuple
the same way as it did for computing the histogram,
and writes the tuple into a buffer corresponding to that partition.
This partitioning routine is based on well-known techniques
using streaming stores and software write-combining%
~\cite{Polychroniou2014,Schuhknecht2015,Muller2015}
to achieve the full memory bandwidth.
When the buffer of a partition is full,
it is sent to the target rank using an asynchronous RDMA write operation,
it replaces the buffer with an empty one,
and continues partitioning the input immediately.
This overlaps computation with communication
and increases performance.

In the network partitioning phase, as in the original algorithm,
each rank compresses the 16-byte workload of the algorithm into 8 bytes
to reduce the data transmitted by a factor of two. This optimization comes as an
additional pass to our query compiler, and although it is very specific, it is
useful for dictionary-encoded data.
The compression uses the fact that some bits of the key
are common for each partition. Specifically,
if we use the identity hash function
and radix partitioning with a fan-out of $2^{\text{F}}$,
the first $\text{F}$ bits of each partition are identical.
Furthermore, we assume that keys and values come from a dense domain
and can be represented with $\text{P}$ bits each.
Thus, key and value can be stored in a single 64-bit word
if $2\cdot \text{P} - \text{F} \leq 64$. After the partitioning and compression,
the operator returns the partitions as
$\langle\varname{networkPartitionID},\varname{partitionData}\rangle$ pairs
such that all tuples of each partition end up on only one rank.
Because we extend the original algorithm with materialization of the input
tuples, we recover the missing bits by forwarding the
$\varname{networkPartitionID}$ further downstream. As we can observe, the
\texttt{MPIExchange} operator batches the tuples in order to avoid
the overhead of sending a tuple-at-a-time over the network.

The subsequent plan joins
the tuples inside two corresponding partitions of the two sides.
An imperative implementation would express this
as a loop over matching partition pairs.
In Modularis, we use the \texttt{NestedMap} operator for the same purpose:
\begin{revision}We take the corresponding
$\langle\varname{networkPartitionID},\varname{partitionData}\rangle$ pairs
and pass them through a \texttt{Zip} operator, which produces
$\langle\varname{networkPartitionID}$, \\ $\varname{partitionData},
 \varname{networkPartitionID}$, $\varname{partitionData}\rangle$ tuples
(note that they are produced in dense, ordered sequence).\end{revision}
This way, all data belonging to one partition pair
is represented in a single tuple,
and we can express the remaining logic as a nested plan
transforming each such tuple.
The nested plan starts by dissecting the input tuple.
The tuple has four fields:
the partition ID and data of the two sides, respectively.
A sequence of \texttt{ParameterLookup} (which returns the entire tuple)
and \texttt{Projection} operators (which retains one of the fields)
extracts one of the fields, each.
The partition data is partitioned further on both sides
by a sequence of \texttt{RowScan}
(which extracts individual tuples from the nested collection
inside the $\varname{partitionData}$ fields),
\texttt{LocalHistogram}, and \texttt{LocalPartitioning} operators.
Note that each of these sequences returns several
$\langle\varname{localPartitionID},\varname{partitionData}\rangle$ pairs.
To be able to recover the dropped bits further downstream,
we augment each of these pairs with the $\varname{networkPartitionID}$
by using the \texttt{CartesianProduct} operator.
Its left side only consists of a single tuple
(containing the network partition ID),
so it does not increase the number of tuples.

The hash and probe phase happens inside another nested plan,
which is executed for each pair of sub-partitions.
As before, we use a \texttt{Zip} operator
to combine all information of each pair of partitions into a single tuple,
on which we call a nested plan using \texttt{NestedMap}.
We use sequences of \texttt{ParameterLookup} and \texttt{Projection} operators
to extract the partitions of the two sides.
Each partition is read by a \texttt{RowScan} operator
and individual tuples produced by these two
are finally fed into the \texttt{BuildProbe} operator,
which produces the matching pairs.
To recover the dropped bits from the network phase,
we use a \texttt{ParametrizedMap} operator:
It contains a function that,
given a parameter from upstream (the network partition ID),
shifts that parameter by a certain amount
and adds the result to the key field
of each input tuple from the other upstream.

The remainder of the plan depends on what happens with the join output.
The Figure shows a plan that materializes that result.
Since each \texttt{NestedMap} needs to return a single tuple,
the result of each nested plan needs to be materialized
using a \texttt{MaterializeRowVector} operator.
This operator produces a single tuple
containing its input tuples as a nested $\varname{RowVector}$.
Each \texttt{NestedMap} thus returns several such tuples
(one for each input tuples)
and the inner tuples can be recovered as a flat stream
using \texttt{RowScan} operators.

In the above description, we reuse many of our building blocks
to construct the high-performance distributed join. Also, we showcase many of
our design principles in action, such as the dedicated read/write operators to
read data either from RMA windows or local partitions, how the high-level
control functions work, and finally how we reuse operators across different
phases of the algorithm.
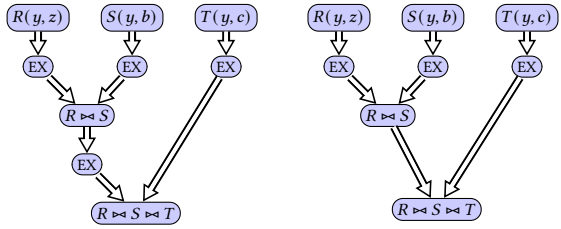
\begin{figure}
\scriptsize
\centering
   \begin{tikzpicture}[node distance=1.25cm, auto]
       \node [smallblock] (first) {$R(y, z)$};
       \node [smallblock, right of=first] (second) {$S(y, b)$};
       \node [smallblock, right of=second] (third) {$T(y, c)$};
       \node [smallblock, below of=first, node distance=0.65cm] (nw1) {EX};
       \node [smallblock, below of=second, node distance=0.65cm] (nw2) {EX};
       \node [smallblock, below of=third, node distance=0.65cm] (nw3) {EX};
       \node [smallblock, below of=nw1, node distance=0.65cm, xshift=0.65cm] (lp1) {$R\bowtie S$};
       \node [smallblock, below of=lp1, node distance=0.65cm] (nw4) {EX};
       \node [smallblock, below of=nw4, node distance=0.65cm, xshift=0.65cm] (lp2) {$R\bowtie S \bowtie T$};

       \draw [vecArrow] (first) -- (nw1);
       \draw [vecArrow] (second) -- (nw2);
       \draw [vecArrow] (third) -- (nw3);
       \draw [vecArrow] (nw1) -- (lp1);
       \draw [vecArrow] (nw2) -- (lp1);
       \draw [vecArrow] (lp1) -- (nw4);
       \draw [vecArrow] (nw3) -- (lp2);
       \draw [vecArrow] (nw4) -- (lp2);

       \node [smallblock, right of=first, node distance=4cm] (first2) {$R(y, z)$};
       \node [smallblock, right of=first2] (second2) {$S(y, b)$};
       \node [smallblock, right of=second2] (third2) {$T(y, c)$};
       \node [smallblock, below of=first2, node distance=0.65cm] (nw11) {EX};
       \node [smallblock, below of=second2, node distance=0.65cm] (nw22) {EX};
       \node [smallblock, below of=third2, node distance=0.65cm] (nw33) {EX};
       \node [smallblock, below of=nw11, node distance=0.65cm, xshift=0.65cm] (lp11) {$R\bowtie S$};
        \node [smallblock, below of=lp11, node distance=1.25cm, xshift=0.65cm] (lp22) {$R\bowtie S \bowtie T$};

       \draw [vecArrow] (first2) -- (nw11);
       \draw [vecArrow] (second2) -- (nw22);
       \draw [vecArrow] (third2) -- (nw33);
       \draw [vecArrow] (nw11) -- (lp11);
       \draw [vecArrow] (nw22) -- (lp11);
       \draw [vecArrow] (lp11) -- (lp22);
       \draw [vecArrow] (nw33) -- (lp22);
    \end{tikzpicture}
\caption{Naive (left) and optimized (right) versions for a sequence of two joins
on the same attribute}\label{fig:complex_sequences}

\vspace{-4mm}
\end{figure}
\subsection{Sequences of joins}
A key advantage of Modularis is that, once we have the original join algorithm,
it is straightforward to extend the plan to run sequences of
joins. None of the work to date on high-performance joins on \mbox{multi-core} CPUs or over
RDMA has ever addressed this design due to the complexity of modifying
the highly tuned operators.
For a cascade of $N$ joins, the output of the
$n-1$-th join is joined with the $n$-th
relation, where $n = 1, \dots, N$. Therefore, in the original plan of
Figure~\ref{fig:modular_join}, after the \texttt{RowScan} operator, we
return the new data to the \texttt{LocalHistogram} and
\texttt{MpiExchange} operators. On the other side,
another upstream operator returns tuples that go through the network
partitioning phase. This pattern is repeated on one side of the corresponding
join for each output and on the other side for the corresponding new relation,
until all of the $N$ joins are performed.

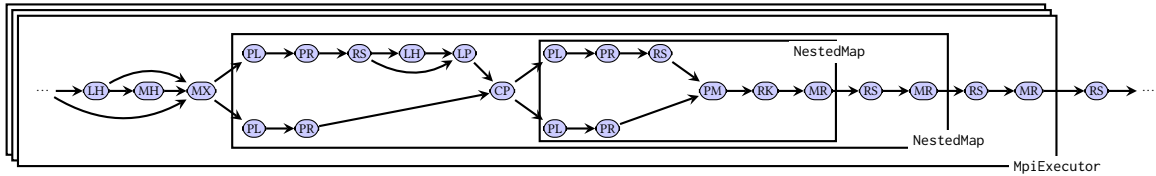
\begin{figure*}[b]
    \centering
    \tiny
    \begin{tikzpicture}[node distance=0.7cm, auto]
    \node [rectangle, thick] (dots1) {\dots};
    \node [smallblock, right of=dots1] (lh1) {LH};
    \node [smallblock, right of=lh1] (gh1) {MH};
    \node [smallblock, right of=gh1] (nw1) {MX};
    \node [smallblock, right of=nw1, yshift=0.5cm] (pl3) {PL};
    \node [smallblock, right of=nw1, yshift=-0.5cm] (pl4) {PL};
    \node [smallblock, right of=pl3] (pr1) {PR};
    \node [smallblock, right of=pl4] (pr2) {PR};
    \node [smallblock, right of=pr1] (rs1) {RS};
    \node [smallblock, right of=rs1] (lh2) {LH};
    \node [smallblock, right of=lh2] (lp1) {LP};
    \node [smallblock, below of=lp1, node distance=0.5cm, xshift=0.5cm] (cp1) {CP};
    \node [smallblock, right of=cp1, yshift=0.5cm] (pl7) {PL};
    \node [smallblock, right of=cp1, yshift=-0.5cm] (pl8) {PL};
    \node [smallblock, right of=pl7] (pr6) {PR};
    \node [smallblock, right of=pl8] (pr7) {PR};
    \node [smallblock, right of=pr6] (rs3) {RS};
    \node [smallblock, right of=rs3, yshift=-0.5cm] (pm) {PM};
    \node [smallblock, right of=pm] (rk) {RK};
    \node [smallblock, right of=rk] (mrv1) {MR};
    \node [smallblock, right of=mrv1] (rs5) {RS};
    \node [smallblock, right of=rs5] (mrv2) {MR};
    \node [smallblock, right of=mrv2] (rs6) {RS};
    \node [smallblock, right of=rs6] (mrv3) {MR};
    \begin{scope}[on background layer]
    \node [cascaded, inner xsep=1em, inner ysep=2em, thick, fit=
    (dots1)(nw1)(pl3)(pl4)(pr1)(pr2)(rs1)(lp1)(cp1)(pl7)(pl8)(pr6)(pr7)(rs3)(pm)(rk)(mrv1)(mrv2)(rs5)(rs6)(mrv3)] (inner1) {};
     \node[fill=white] at (inner1.south east) {\scriptsize \texttt{MpiExecutor}};
    \end{scope}
    \node [draw, thick, inner xsep=0.75em, inner ysep=0.75em, fit=
    (pl3)(pl4)(pr1)(pr2)(rs1)(lp1)(cp1)(pl7)(pl8)(pr6)(pr7)(rs3)(pm)(rk)(mrv1)(mrv2)] (inner2) {};
    \node [draw, thick, inner xsep=0.25em, inner ysep=0.25em, fit=
    (pl7)(pl8)(pr6)(pr7)(rs3)(pm)(rk)(mrv1)] (inner3) {};
    \node[fill=white, yshift=0.1cm] at (inner2.south east) {\scriptsize \texttt{NestedMap}};
    \node[fill=white, yshift=-0.17cm, xshift=-0.1cm] at (inner3.north east) {\scriptsize \texttt{NestedMap}};
    \node [smallblock, node distance=0.9cm, right of=mrv3] (frs) {RS};
    \node [rectangle, thick, right of=frs] (fdots) {\dots};

    \draw [arrow] (dots1) -- (lh1);
    \draw [arrow] (dots1) to[bend right] (nw1);
    \draw [arrow] (lh1) -- (gh1);
    \draw [arrow] (lh1) to[bend left] (nw1);
    \draw [arrow] (gh1) -- (nw1);
    \draw [arrow] (nw1) -- (pl3);
    \draw [arrow] (nw1) -- (pl4);
    \draw [arrow] (pl3) -- (pr1);
    \draw [arrow] (pl4) -- (pr2);
    \draw [arrow] (pr1) -- (rs1);
    \draw [arrow] (rs1) -- (lh2);
    \draw [arrow] (rs1) to[bend right] (lp1);
    \draw [arrow] (lh2) -- (lp1);
    \draw [arrow] (pr2) -- (cp1);
    \draw [arrow] (lp1) -- (cp1);
    \draw [arrow] (cp1) -- (pl7);
    \draw [arrow] (cp1) -- (pl8);
    \draw [arrow] (pl7) -- (pr6);
    \draw [arrow] (pl8) -- (pr7);
    \draw [arrow] (pr6) -- (rs3);
    \draw [arrow] (pr7) -- (pm);
    \draw [arrow] (rs3) -- (pm);
    \draw [arrow] (pm) -- (rk);
    \draw [arrow] (rk) -- (mrv1);
    \draw [arrow] (mrv1) -- (rs5);
    \draw [arrow] (rs5) -- (mrv2);
    \draw [arrow] (mrv2) -- (rs6);
    \draw [arrow] (rs6) -- (mrv3);
    \draw [arrow] (mrv3) -- (frs);
    \draw [arrow] (frs) -- (fdots);
\end{tikzpicture}
    \vspace{-2.5mm}
    \caption{Plan that runs the distributed \texttt{GROUP BY} with modular operators across many nodes}\label{fig:modular_groupby}
    \vspace{-1mm}
\end{figure*}
However, if all the joins are on the same attribute \begin{revision} (i.e., the
    attribute $y$ in Figure~\ref{fig:complex_sequences})
\end{revision} and the relations fit in
main memory, we can apply the following optimization: We
network-partition all relations at the beginning instead of reshuffling
the output of every join through the network. \begin{revision} This is possible
since we execute the output of the first join again on the attribute $y$,
and therefore we can pre-partition all the relations from the beginning of
the query instead of waiting for the result of each join \end{revision}.
This way, for a cascade of
$N$ joins, we shuffle through the network $N+1$ instead of $2N$ relations.
We show our optimization for a sequence of two joins in
Figure~\ref{fig:complex_sequences}, where instead of shuffling four
relations through the network, namely $R$, $S$, $T$, and the output of $R\bowtie S$,
we shuffle only $R$, $S$, and $T$. \begin{revision}For conciseness, we depict
with the operator \texttt{EX} the chain of a LocalHistogram, MPI
Histogram, and MPI Exchange operators.\end{revision}

This optimization is easily applied because of the operator modularity.
In the case of monolithic operators,
a system engineer would have to take special care of
this case by adapting a large part of the system and possibly by reimplementing
parts of the algorithm. In contrast, Modularis restructures the
sub-operators inside the query plans and
takes advantage of the
common attribute in a sequence of joins. After it performs all the
network partitioning phases at the beginning of the inner plan of
Figure~\ref{fig:modular_join}, it carries out all the local
partitioning phases in the first nested map. Finally, it forms a
sequence of \texttt{BuildProbe} operators where the output of the $n-1$-th
\texttt{BuildProbe} is the input of the $n$-th \texttt{BuildProbe}
and the final build probe output is the input of the
\texttt{ParametrizedMap} operator.
\subsection{Distributed \texttt{\large GROUP BY}}\label{sec:complex-plans:group-by}
To illustrate how Modularis simplifies operator development and provides
extensibility, we implement a distributed
\texttt{GROUP BY} operator by re-using components from the previous use cases.
We show the corresponding plan in Figure~\ref{fig:modular_groupby}. The
algorithm workload is a 16-byte tuple (8 bytes for the key and 8 bytes
for the value).

The plan starts with any upstream operator that returns tuples to the
\texttt{LocalHistogram} and \texttt{MpiExchange} operators. Since we
have multiple consumers from one operator, these tuples have to be materialized
and put into a separate pipeline
(the materialization is not shown in the Figure as discussed earlier).
After the local and global histogram
calculations, the tuples are partitioned and distributed through the network.
The operator performs a similar compression scheme as in
Section~\ref{sec:complex-plans:partitioned-hash-join}. As before, this
compression allows us to reduce the network traffic in
half, which is crucial for performance. Every output tuple
(which consists of $\langle\varname{networkPartitionID},\varname{partitionData}\rangle$ pairs)
coming from the \texttt{MpiExchange} operator
is the input of a \texttt{NestedMap}, which executes its nested plan for every
input partition.

The execution of the nested plan starts with the
\texttt{ParameterLookup} operators. The tuples returned by these operators are
passed to \texttt{Projection} operators, which ensure that each downstream
operator gets the correct input. Specifically,
the network partitioning data are passed to a \texttt{RowScan} operator that
returns a tuple at a time to the \texttt{LocalHistogram} and
\texttt{LocalPartitioning} operators. After the histogram calculation,
the \texttt{LocalPartitioning} consumes both the input data and the
calculated histogram to calculate the necessary prefixes inside a partition. It
then performs the data partitioning. The corresponding partitioned data
is concatenated with the network bits that the \texttt{MpiExchange}
removed, using a \texttt{CartesianProduct}. Lastly, the
$\langle\varname{networkPartitionID}, \varname{localPartitionID}$, \\ $\varname{partitionData}\rangle$
tri\-ples are the input to \texttt{NestedMap} operator, which executes
the final aggregation for each input partition.

To perform the final aggregation, we first have to restore the original
keys as we did in the join algorithm after the hash build
and probe phase. The difference now is that we have to restore the full
keys using the \texttt{ParametrizedMap} operator before we forward each tuple
to a \texttt{ReduceByKey} operator, which aggregates the data per local partition.
Afterward, we materialize the output tuples of the \texttt{ReduceByKey} operator
with a \texttt{MaterializeRowVector} operator. Like in the distributed hash join
case, we finish the plan with the \texttt{RowScan} operators that remove nesting
levels of the \texttt{MaterializeRowVector} operators that should end every
nested plan. Finally, the individual
results from the workers return to the driver.

Based on the previous description, it is evident that the distributed
\texttt{GROUP BY} plan is very similar to the distributed
hash join plan. The main differences are 1) the total number of
input relations and 2) that for distributed \texttt{GROUP BY} operator we do not
perform a hash build and probe phase in the end but an
aggregation using a \texttt{ReduceByKey} operator.
The large overlap between the two plans shows how Modularis uses a
similar set of sub-operators to implement different relational database
operators in contrast to using monolithic operators, where the different
operators would probably have to be reimplemented almost from scratch, although
the logic behind shares many similarities (e.g., the partitioning phases).

\begin{figure*}[t]
    \centering
    \tiny
    \begin{tikzpicture}[node distance=0.7cm, auto]
    \node [smallblock] (pl1) {PL};
    \node [smallblock, below of=pl1] (pl2) {PL};
    \node [smallblock, right of=pl1, yshift=-0.375cm] (zip1) {ZP};
    \node [smallblock, right of=pl1, xshift=0.7cm] (pl3) {PL};
    \node [smallblock, right of=pl2, xshift=0.7cm] (pl4) {PL};
    \node [smallblock, right of=pl3] (pr1) {PR};
    \node [smallblock, right of=pl4] (pr2) {PR};
    \node [smallblock, right of=pr1] (ps1) {PS};
    \node [smallblock, right of=pr2] (ps2) {PS};
    \node [smallblock, right of=ps1] (cs1) {CS};
    \node [smallblock, right of=ps2] (cs2) {CS};
    \node [smallblock, right of=cs1] (f1) {F};
    \node [smallblock, right of=cs2] (m1) {M};
    \node [smallblock, right of=f1] (m2) {M};
    \node [smallblock, right of=m1, xshift=0.7cm] (lh1) {LH};
    \node [smallblock, right of=m2] (lh2) {LH};
    \node [smallblock, right of=lh1] (mh1) {MH};
    \node [smallblock, right of=lh2] (mh2) {MH};
    \node [smallblock, right of=mh1] (mx1) {MX};
    \node [smallblock, right of=mh2] (mx2) {MX};
    \node [smallblock, right of=mx1] (rs3) {RS};
    \node [smallblock, right of=mx2] (rs4) {RS};
    \node [smallblock, right of=rs4, yshift=-0.375cm] (hp) {HP};
    \node [smallblock, right of=hp] (m3) {M};
    \node [smallblock, right of=m3] (m4) {M};
    \node [smallblock, right of=m4] (rk1) {RK};
    \node [smallblock, right of=rk1] (mp) {MP};
    \node [smallblock, right of=mp] (ps5) {PS};
    \node [smallblock, right of=ps5] (cs5) {CS};
    \node [smallblock, right of=cs5] (rk2) {RK};
    \node [smallblock, right of=rk2] (tk) {TK};
    \node [smallblock, right of=tk] (mr) {MR};
    \begin{scope}[on background layer]
    \node [cascaded, inner xsep=1em, inner ysep=2em, fit=
    (pl3)(pl4)(pr1)(pr2)(ps1)(ps2)(cs1)(cs2)(f1)(m1)(m2)
    (lh1)(lh2)(mh1)(mh2)(mx1)(mx2)(rs3)(rs4)(hp)(m3)(m4)(rk1)(mp)] (inner1) {};
    \node[fill=white] at (inner1.south east) {\scriptsize \texttt{MpiExecutor}};
    \end{scope}
    \draw [arrow] (pl1) -- (zip1);
    \draw [arrow] (pl2) -- (zip1);
    \draw [arrow] (zip1) -- (pl3);
    \draw [arrow] (zip1) -- (pl4);
    \draw [arrow] (pl3) -- (pr1);
    \draw [arrow] (pl4) -- (pr2);
    \draw [arrow] (pr1) -- (ps1);
    \draw [arrow] (pr2) -- (ps2);
    \draw [arrow] (ps1) -- (cs1);
    \draw [arrow] (ps2) -- (cs2);
    \draw [arrow] (cs1) -- (f1);
    \draw [arrow] (cs2) -- (m1);
    \draw [arrow] (f1) -- (m2);
    \draw [arrow] (m1) -- (lh1);
    \draw [arrow] (m2) -- (lh2);
    \draw [arrow] (lh1) -- (mh1);
    \draw [arrow] (lh1) to[bend left] (mx1);
    \draw [arrow] (lh2) -- (mh2);
    \draw [arrow] (m1) to[bend right] (mx1);
    \draw [arrow] (lh2) to[bend left] (mx2);
    \draw [arrow] (m2) to[bend right] (mx2);
    \draw [arrow] (mh1) -- (mx1);
    \draw [arrow] (mh2) -- (mx2);
    \draw [arrow] (mx1) -- (rs3);
    \draw [arrow] (mx2) -- (rs4);
    \draw [arrow] (rs3) -- (hp);
    \draw [arrow] (rs4) -- (hp);
    \draw [arrow] (hp) -- (m3);
    \draw [arrow] (m3) -- (m4);
    \draw [arrow] (m4) -- (rk1);
    \draw [arrow] (rk1) -- (mp);
    \draw [arrow] (mp) -- (ps5);
    \draw [arrow] (ps5) -- (cs5);
    \draw [arrow] (cs5) -- (rk2);
    \draw [arrow] (rk2) -- (tk);
    \draw [arrow] (tk) -- (mr);
\end{tikzpicture}
    \vspace{-2.5mm}
    \caption{Modularis plan for TPC-H Q12 on RDMA}\label{fig:tpch_rdma}
    \vspace{-2mm}
\end{figure*}
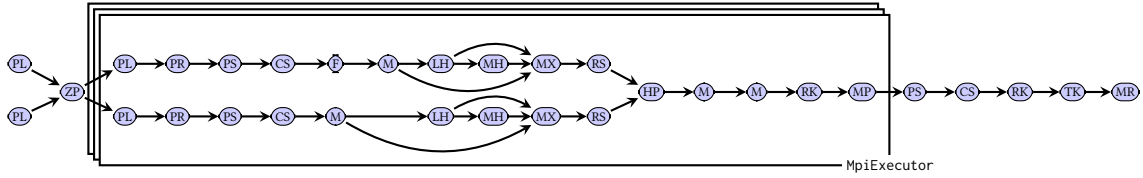

\begin{figure*}[t]
    \centering
    \tiny
    \begin{tikzpicture}[node distance=0.7cm, auto]
    \node [smallblock] (pl1) {PL};
    \node [smallblock, below of=pl1] (pl2) {PL};
    \node [smallblock, right of=pl1, yshift=-0.375cm] (zip1) {ZP};
    \node [smallblock, right of=pl1, xshift=0.7cm] (pl3) {PL};
    \node [smallblock, right of=pl2, xshift=0.7cm] (pl4) {PL};
    \node [smallblock, right of=pl3] (pr1) {PR};
    \node [smallblock, right of=pl4] (pr2) {PR};
    \node [smallblock, right of=pr1] (ps1) {PS};
    \node [smallblock, right of=pr2] (ps2) {PS};
    \node [smallblock, right of=ps1] (cs1) {CS};
    \node [smallblock, right of=ps2] (cs2) {CS};
    \node [smallblock, right of=cs1] (f1) {F};
    \node [smallblock, right of=cs2] (m1) {M};
    \node [smallblock, right of=f1] (m2) {M};
    \node [smallblock, right of=m1, xshift=0.7cm] (par1) {PAR};
    \node [smallblock, right of=m2] (par2) {PAR};
    \node [smallblock, right of=par1] (rs1) {RS};
    \node [smallblock, right of=par2] (rs2) {RS};
    \node [smallblock, right of=rs1] (s3x1) {S3X};
    \node [smallblock, right of=rs2] (s3x2) {S3X};
    \node [smallblock, right of=s3x1] (ps3) {PS};
    \node [smallblock, right of=s3x2] (ps4) {PS};
    \node [smallblock, right of=ps3] (cs3) {CS};
    \node [smallblock, right of=ps4] (cs4) {CS};
    \node [smallblock, right of=cs4, yshift=-0.375cm] (hp) {HP};
    \node [smallblock, right of=hp] (m3) {M};
    \node [smallblock, right of=m3] (m4) {M};
    \node [smallblock, right of=m4] (rk1) {RK};
    \node [smallblock, right of=rk1] (mp) {MP};
    \node [smallblock, right of=mp] (ps5) {PS};
    \node [smallblock, right of=ps5] (cs5) {CS};
    \node [smallblock, right of=cs5] (rk2) {RK};
    \node [smallblock, right of=rk2] (tk) {TK};
    \node [smallblock, right of=tk] (mr) {MR};
    \begin{scope}[on background layer]
    \node [cascaded, inner xsep=1em, inner ysep=2em, fit=
    (pl3)(pl4)(pr1)(pr2)(ps1)(ps2)(cs1)(cs2)(f1)(m1)(m2)
    (par1)(par2)(rs1)(rs2)(s3x1)(s3x2)(ps3)(ps4)(cs3)(cs4)(hp)(m3)(m4)(rk1)(mp)] (inner1) {};
    \node[fill=white] at (inner1.south east) {\scriptsize \texttt{LambdaExecutor}};
    \end{scope}
    \draw [arrow] (pl1) -- (zip1);
    \draw [arrow] (pl2) -- (zip1);
    \draw [arrow] (zip1) -- (pl3);
    \draw [arrow] (zip1) -- (pl4);
    \draw [arrow] (pl3) -- (pr1);
    \draw [arrow] (pl4) -- (pr2);
    \draw [arrow] (pr1) -- (ps1);
    \draw [arrow] (pr2) -- (ps2);
    \draw [arrow] (ps1) -- (cs1);
    \draw [arrow] (ps2) -- (cs2);
    \draw [arrow] (cs1) -- (f1);
    \draw [arrow] (cs2) -- (m1);
    \draw [arrow] (f1) -- (m2);
    \draw [arrow] (m1) -- (par1);
    \draw [arrow] (m2) -- (par2);
    \draw [arrow] (par1) -- (rs1);
    \draw [arrow] (par2) -- (rs2);
    \draw [arrow] (rs1) -- (s3x1);
    \draw [arrow] (rs2) -- (s3x2);
    \draw [arrow] (s3x1) -- (ps3);
    \draw [arrow] (s3x2) -- (ps4);
    \draw [arrow] (ps3) -- (cs3);
    \draw [arrow] (ps4) -- (cs4);
    \draw [arrow] (cs3) -- (hp);
    \draw [arrow] (cs4) -- (hp);
    \draw [arrow] (hp) -- (m3);
    \draw [arrow] (m3) -- (m4);
    \draw [arrow] (m4) -- (rk1);
    \draw [arrow] (rk1) -- (mp);
    \draw [arrow] (mp) -- (ps5);
    \draw [arrow] (ps5) -- (cs5);
    \draw [arrow] (cs5) -- (rk2);
    \draw [arrow] (rk2) -- (tk);
    \draw [arrow] (tk) -- (mr);
\end{tikzpicture}
    \vspace{-2.5mm}
    \caption{Modularis plan for TPC-H Q12 on Serverless}\label{fig:tpch_lambda}
    \vspace{-2.5mm}
\end{figure*}
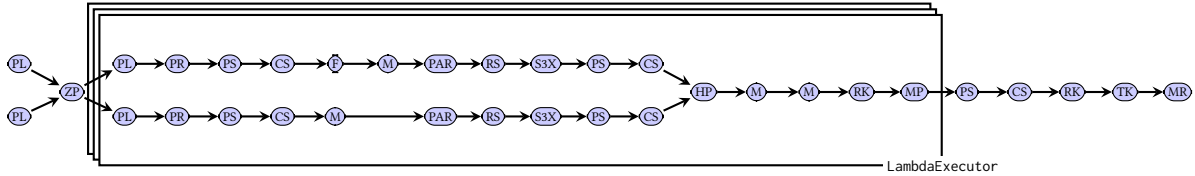

\subsection{TPC-H queries}
\label{subsec:tpch-queries}
So far we have used Modularis to implement key components of a relational
query processor. We can use the same sub-operators to implement TPC-H queries.
In fact, by altering only our MPI-specific sub-operators to
Lambda-specific we can execute the same set of TPC-H queries on a different
hardware platform. The latter is a strong argument towards modularity, as
systems that operate on different hardware platforms have fundamentally different
execution layers making it almost impossible to apply the techniques used
in one for the other without major reimplementation.

We take as an example TPC-H Q12 and show (simplified)
query plans of Modularis in Figures~\ref{fig:tpch_rdma}
and~\ref{fig:tpch_lambda} for RDMA and serverless
respectively. Although the plans are simplified, they still
preserve the semantics of the system.
Since the network bandwidth in serverless is rather slow
(around \SI{80}{\mega\bit\per\second}, see \cite{muller2020lambada}),
we use the partitioning only as a pre-processing step
required by the exchange operator
and do not partition the exchanged data further.
This means that each worker
processes only one data partition after the exchange,
so we do not have any \texttt{NestedMap} operators in that platform.
Both plans take Parquet file paths to the base tables as input,
either in S3 or NFS depending on the platform.
The paths of the left and right input are zipped
and each resulting pair is used as the input
for a nested plan instance by the executor of the respective platform.
The execution of a nested plan starts again
with \texttt{ParameterLookup} operators,
from which we project the paths for the left and right relations,
pass them to \texttt{ParquetScan} operators,
and finally extract individual tuples from the column chunks
produced by that operator using the \texttt{ColumnScan} operator.
The next operators express the corresponding operations of Query 12.

At this point, we differentiate the plans with the operators that constitute the
exchange routine for each platform. Note that we do not perform any compression
of the tuples, such as in the case of the distributed join and the \texttt{GROUP
BY}. In the case of RDMA, the plan looks similar
to the ones that we have presented before. In the case of serverless, we use the
exchange algorithm of Lambada~\cite{muller2020lambada}: First, we
partition the data into a sequence of partitions using a \texttt{Partition}
operator. Subsequently, the \texttt{GroupBy} operator takes the
\texttt{<pid, data>} partitions and groups them by \texttt{pid}. Then the
\texttt{RowScan} operator reads each partition and forwards it to the
\texttt{S3Exchange} operator, which writes the data into a file on S3 whose file
name is based on the ID of the sender containing one row group per receiver.
The \texttt{S3Exchange} then returns triples of worker-specific S3 paths and the first and last row group to read
from that file, which is then read by the subsequent \texttt{ParquetScan} operator. This pattern implements the ``write combining'' optimization of Lambada, which significantly reduces the number of write requests to S3.
Finally, the column chunks produced by the \texttt{ParquetScan} operator
are consumed by a \texttt{ColumnScan}, which extracts individual tuples.

After the exchange phase finishes, the plans converge again: We perform
the join of the two relations using the \texttt{HashProbe} operator. The
matched tuples are transformed with a series of \texttt{Map} and
\texttt{ReduceByKey} operators to get
the final result, and we use a \texttt{MaterializeParquet} operator to return
the individual results of the workers to the driver process. We read
the individual results using a sequence of \texttt{ParquetScan} and
\texttt{ColumnScan} operators. Finally, we merge the results of the workers using a
\texttt{ReduceByKey} operator, we select the first tuple using a \texttt{TopK}
operator and we return the results to the user with a
\texttt{MaterializeRowVector} operator.

We therefore show how altering only a handful of operators that are
hardware-specific, allows us to run the same TPC-H query on two very different
hardware platforms. That way, implementation effort is reduced vastly. Using the
same ideas, we could extend the TPC-H implementation to use an exchange
operator based on TCP. The addition of more backends only requires changing the
executor and the operators that comprise the network exchange phase.

\vspace{-0.75mm}
\begin{revision}
\subsection{Integration with smart Storage}
\label{subsec:smart_storage}
To show how Modularis can use a smart
storage component offered by a major cloud provider, we integrate
S3Select~\cite{s3select} into our system. S3Select is a
smart storage engine offered by Amazon
that follows the trend of pushing computation into storage~\cite{woods2014ibex, jo2016yoursql}
to overcome the I/O bottleneck.
S3Select pushes computations directly into
S3 and thus it pulls only the data that the user needs from these
objects.

S3Select takes as parameters an SQL query, the S3 input path, and an output
serialization format (either JSON or CSV). In our case, the user writes an SQL
expression in the frontend, which pushes selections and projections
to S3Select. We
have created an additional sub-operator called \texttt{S3SelectScan}
that our query compiler decomposes into three simpler, more re-usable separate operators. The
first sub-operator performs an API call to S3Select and requests
the data, which S3Select returns in CSV format. The sub-operator
then uses Apache Arrow
to convert the CSV to an Arrow Table and forwards the table downstream.
The next operator converts the Arrow Table to a
\textit{collection of tuples} as this is defined in
Section~\ref{subsec:interface}.
Finally, the third operator is a \texttt{ColumnScan} that gets this
collection and returns the individual tuples. Then, we continue with
the execution of the rest of the plan in our serverless backend.

Thus, by following the same design principles, we only develop the sub-operators
needed to integrate S3Select into Modularis. We did not have to redesign the
system from scratch to adapt to this system component. We only had to make small
adjustments, whereas other systems would have to make a major redesign to
incorporate such a change. Finally, this integration shows the generality of our
design principles, because the implementation done is not limited
to S3Select. We could use the same architecture to request data from
a different smart storage engine that is based on an accelerator (like Amazon
AQUA~\cite{aqua}). As databases push more
computation to storage or to accelerators, we expect that systems like Modularis
will be able to use these components with only minimal changes.

\end{revision}
\vspace{-0.5em}

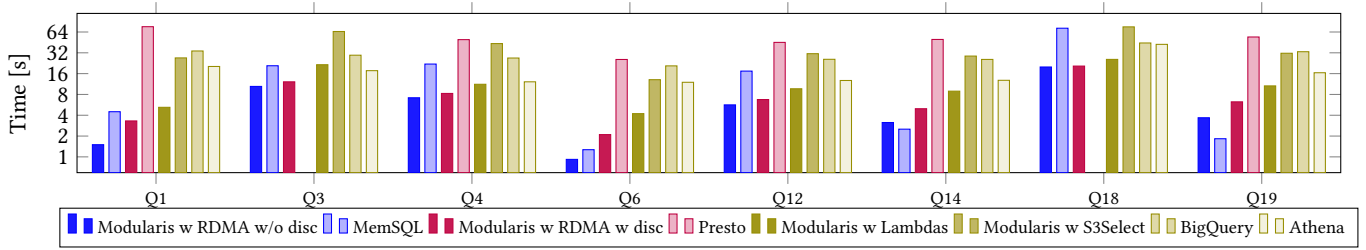
\begin{figure*}[t]
  \centering
  \begin{tikzpicture}
\begin{axis}[
    ybar,
    ymode=log,
    log origin=infty,
    x=2.1cm,
    enlarge x limits={abs=1.05cm},
    ymin=0,
    legend style={at={(0.5,-0.22)},
    anchor=north,legend columns=-1, font=\footnotesize},
    ylabel={Time [s]},
    symbolic x coords={Q1, Q3, Q4, Q6, Q12, Q14, Q18, Q19},
    xtick=data,
    ytick={1, 2, 4, 8, 16, 32, 64},
    log ticks with fixed point,
    legend entries={Modularis w RDMA w/o disc, MemSQL, Modularis w RDMA w disc, Presto,
    Modularis w Lambdas, Modularis w S3Select, BigQuery, Athena},
    bar width=0.15cm,
    nodes near coords align={vertical},
    x tick label style={font=\footnotesize,text width=1cm,align=center},
    ylabel near ticks,
    every node near coord/.append style={font=\small},
    height=3.7cm,
    width=8cm
    ]
    \addplot[color=blue, fill=blue!90!white] coordinates {(Q1, 1.50) (Q3, 10.45) (Q4, 7.16) (Q6, 0.92) (Q12,
            5.65) (Q14, 3.14) (Q18, 19.96) (Q19, 3.68) } ;
    \addplot[color=blue, fill=blue!30!white] coordinates {(Q1, 4.50) (Q3, 20.8) (Q4, 22) (Q6, 1.27) (Q12, 17.4)
            (Q14, 2.516) (Q18, 72.6) (Q19, 1.83) } ;
    \addplot[color=purple, fill=purple!90!white] coordinates {(Q1, 3.31) (Q3, 12.17) (Q4, 8.27) (Q6, 2.10) (Q12,
            6.75) (Q14, 4.97) (Q18, 20.57) (Q19, 6.25) } ;
    \addplot[color=purple, fill=purple!30!white] coordinates {(Q1, 76.39) (Q3, 0) (Q4, 49.63) (Q6, 25.63) (Q12,
            45.28) (Q14, 49.94) (Q18, 0) (Q19, 54.196) } ;
    \addplot[color=olive, fill=olive!90!white]  coordinates {(Q1, 5.21) (Q3,
            21.52) (Q4, 11.22) (Q6, 4.22) (Q12,
            9.66) (Q14, 8.92) (Q18, 25.70) (Q19, 10.63) } ;
    \addplot[color=olive, fill=olive!60!white]  coordinates {(Q1, 27) (Q3, 65.0548) (Q4, 43.524) (Q6, 13.071) (Q12,
            31.0844) (Q14, 28.7088) (Q18, 75.932) (Q19, 31.5336) } ;
    \addplot[color=olive, fill=olive!30!white] coordinates {(Q1, 34.01) (Q3, 29.55) (Q4, 26.90) (Q6, 20.71) (Q12,
            25.78) (Q14, 25.68) (Q18, 44.40) (Q19, 33.20) } ;
    \addplot[color=olive, fill=olive!10!white] coordinates {(Q1, 20.30) (Q3, 17.66) (Q4, 12.18) (Q6, 11.99) (Q12,
            12.73) (Q14, 12.81) (Q18, 42.40) (Q19, 16.48) } ;
\end{axis}
\end{tikzpicture}
  \vspace{-1.5em}
  \caption{TPC-H queries runtime using SF-500}\label{fig:tpch-experiments}
  \vspace{-1em}
\end{figure*}
\section{Modularis evaluation}
In this section, we first compare Modularis to commercial systems on TPC-H
queries on two hardware platforms: RDMA and serverless. We then analyze the
performance of the system against a distributed hash join. Finally, we show
the runtime of a distributed \texttt{GROUP BY} operator and variations of
plans for sequences of joins.
We run all of the RDMA experiments using all available cores from a cluster of 8
machines (specifications in Table~\ref{tab:machinespecs}).
Regarding the MPI
implementation, we use OpenMPI 3.1.4 as opposed to
foMPI~\cite{gerstenberger2014enabling} used by \cite{Barthels2017} because
foMPI is specific to Cray machines. Finally, unless otherwise mentioned,
we run each experiment five times and report the average among such runs.
\begin{table}[b]
    \centering
    \vspace{-2mm}
    \begin{tabular}{c c}
        \toprule
        \textbf{Component} & \textbf{Specs} \\
        \midrule
        CPUs & 2 $\times$ Intel Xeon E5--2609 2.40 GHz \\
        Cores/Threads & 2 $\times$ 4/4 \\
        RAM & 128 GB \\
        L1 Cache & 2 $\times$ 4 $\times$ 64 KB \\
        L2 Cache & 2 $\times$ 4 $\times$ 256 KB \\
        L3 Cache & 2 $\times$ 10 MB \\
        InfiniBand & Mellanox QDR HCA \\
        \bottomrule
    \end{tabular}
    \caption{RDMA cluster specification}\label{tab:machinespecs}
    \vspace{-2mm}
\end{table}
\subsection{TPC-H queries}

We start our evaluation by comparing Modularis to commercial systems, both for
RDMA and serverless using TPC-H queries. We present the results for
scale factor $500$ in Figure~\ref{fig:tpch-experiments}.
We pick TPC-H Queries~1, 3, 4, 6, 12, 14, 18, and 19, which is more than a
third of the benchmark and a representative subset of all the challenges that the
benchmark poses. More specifically, Q1 has a large aggregation, Q3 has a large
join, Q4 involves an inclusion test, Q6, Q14, and Q19 have selective
filtering that largely reduces the processed tuples of the input tables, Q12
involves almost all of the major operators a relational engine should implement
(join, selection, projection, aggregation, sorting), and Q18 has a
high-cardinality aggregation. As mentioned, the challenges involved in
these queries together make up for the most important challenges of the benchmark itself and an
execution layer that has a good performance on these queries has a high
probability to perform well on the whole
benchmark~\cite{kersten2018everything, boncz2013tpc}.
The inclusion of more queries is not a limitation of the system
but  involves the implementation of a more sophisticated
optimizer, which is part of future work and out of the scope of this paper;
for now, we concentrate on the execution layer alone.

\subsubsection{RDMA}
For the RDMA backend, we compare Modularis against two different
systems, Presto and SingleStore (previously MemSQL). Both of these systems do not use
RDMA for the network exchange but we configure them to use the InfiniBand
network for data transfer at higher rates. We verify this
by monitoring the network traffic. For all the RDMA experiments, we run
Modularis using 64 workers.
For SingleStore, we use version 7.0.12 and deploy a master aggregator node and 7
leaf nodes. We do a warm run for each of the queries and report averages
of 5 runs. We deploy the TPC-H database using both a
row-store~\cite{memsqlrow} and a column-store~\cite{memsqlcolumn}
format. The row-store format is completely in-memory and has very good random
seek performance. The disk-backed
column-store format has optimizations such as indices, compression, fast
aggregations, and table scans, most of which are not supported by Modularis.
We use the column-store format, as it is the best for all the
queries. Although the column-store format is disk-backed, SingleStore serves
all queries from an in-memory cache; for a fair comparison, we thus exclude
the time Modularis needs to read the data from disk as well.
We observe that Modularis is between
30\% up to more than 3x faster for all the queries except Q14 and Q19.

For Q19, Modularis is slower because
the histogram-based RDMA network exchange is slower than a
broadcast operator for small joins.  Modularis is 40\%
slower in Q14, because the \texttt{Map} operator does not
perform the selective \texttt{LIKE} as fast as SingleStore. For the queries
where Modularis is faster, all except Queries 1 and 6 have large joins,
where the RDMA network exchange has better performance. Another operator
responsible for a large part of the overall runtime is \texttt{ReduceByKey},
which uses a highly optimized version of a parallel hash map in Modularis and
thus executes very fast large aggregations. The difference is obvious
in Queries 1 and 18 where Modularis is 3 and 3.5 times faster,
respectively.

For Presto, we deploy Presto SQL (now Trino) version 327 along with HDFS (Hadoop version 2.6.0) in the
eight-machine cluster using one node exclusively as coordinator and NameNode.
We configure HDFS to use replication factor 3 and Presto to use as much memory
as possible. We run each query four times, use the first run as a warm-up and
then report the average of the other runs. To have a fair
comparison, we include also the time that Modularis needs to read the input
data from disk. For Queries
3 and 18, Presto cannot execute the queries and fails because of insufficient
resources. Our system is 6-9x faster than Presto, depending on the query, partly
because of the optimized RDMA network exchange and partly because of the
highly-performance sub-operators.

\subsubsection{Serverless}
For the serverless backend, we use the two Query-as-a-Service systems
Athena and BigQuery as baselines.
These systems run queries over cloud storage without the
need to start or maintain any infrastructure.
For BigQuery, we use external tables~\cite{bigquery}
that point to 512 Parquet files per relation stored on a single-region bucket
in Google Cloud
Storage in the standard storage tier. Creating such a table is just a metadata
operation and takes <1s, i.e., no data is loaded. Instead, the query runs
against the original files on cloud storage. We use the same files for Athena,
for which we created an external table~\cite{athena} as well.
The remaining configuration is done by the cloud provider.
For Modularis we use the same Parquet files. \begin{revision}
For the integration with S3Select,
we use 512 workers and for the experiments with the
\texttt{ParquetScanOperator} we use
256 workers. The workers for both these configurations have \SI{2}{\gibi\byte}
of main memory each.
For all three systems we run each query 6 times and report averages.
As we observe, Modularis with S3Select is comparable to
BigQuery for the majority of the queries but slower than Athena.
To investigate the reason, we isolate the calls to S3Select and
time them independently of Modularis. We find out that they make up for the
majority of the runtime (e.g. for Q1, 24 out of 27 seconds are spent getting data
from S3Select). The problem is enlarged when we read many relations (e.g. Q3,
Q18) and is almost negligible for high selective queries that read only one
relation (e.g. Q6). In fact, for Q6 Modularis with S3Select is very close to
Athena and faster than BigQuery. The larger running times of calls to S3Select
are because the service returns chunks of uncompressed CSV data,
whereas our \texttt{ParquetScan} reads data in compressed format and
also pushes down projections.
Our observations agree with the ones made in~\cite{yu2020pushdowndb}.
This problem is ameliorated if
we read only one Parquet file per worker, which is why
we use 512 workers in this experiment.

We can improve the performance significantly by using the specialized
\texttt{ParquetScan} operator. In that case, Modularis outperforms both baselines
in all queries except for Query 3, where Athena is marginally faster
thanks to a better query plan.
For Queries 1 and 6, the
running time difference is due to the \texttt{ParquetScan} operator, which
is optimized to push down projections while reading data in
compressed format. For the other queries, the specialized
\texttt{LambdaExchange} operator can run workloads involving data exchange
between workers faster than the commercial baselines. This is evident
because the workers are mostly bounded by network
bandwidth and latency.
The current performance of S3Select illustrates the advantage of
modularity: today, we can use the much faster alternative of our optimized
\texttt{ParquetScan}, but using \texttt{S3SelectScan} can be enabled easily
if AWS improves the performance of their smart storage engine in the future.
\end{revision}

\begin{figure*}[t]
\begin{subfigure}[t]{0.66\textwidth}
\centering
\begin{tikzpicture}
\begin{axis}[
    name=ax1,
    ybar stacked,
    x=1.2cm,
    enlarge x limits={abs=0.6cm},
    ymin=0,
	bar width=10pt,
    legend style={at={(1.3, 1.3), font=\footnotesize},
    anchor=north,legend columns=-1},
    legend cell align={left},
    ylabel={Time[s]},
    ymin=0,
    symbolic x coords={original, model, modularis},
    xtick=data,
    x tick label style={font=\scriptsize},
    ytick={0, 3, 6, 9},
    ylabel near ticks,
    width=4cm,
    height=3.5cm
    ]
\addplot+[ybar] plot coordinates {(original, 0.3258536) (model, 0.2165328) (modularis, 0.248647)};
\addplot+[ybar] plot coordinates {(original, 0.0023502) (model, 0.0122686375) (modularis, 0.0352534)};
\addplot+[ybar] plot coordinates {(original, 4.8936222) (model, 5.218052563) (modularis, 5.6468926)};
\addplot+[ybar] plot coordinates {(original, 0.8452276) (model, 1.206004) (modularis, 1.9923626)};
\addplot+[ybar] plot coordinates {(original, 1.649278)  (model, 0.6265502) (modularis, 1.3892978)};
\legend{\strut local histogram, \strut global histogram, \strut network
partitioning, \strut local partitioning, \strut build probe}
\end{axis}
\begin{axis}[
    at={(ax1.south east)},
    xshift=2cm,
    name=ax2,
    ybar stacked,
    x=1.2cm,
    enlarge x limits={abs=0.6cm},
    ymin=0,
	bar width=10pt,
    legend style={at={(0.5,-0.10)},
      anchor=north,legend columns=-1},
    legend cell align={left},
    ylabel={Time[s]},
    ymin=0,
    symbolic x coords={original, model, modularis},
    xtick=data,
    x tick label style={font=\scriptsize},
    ylabel near ticks,
    width=4cm,
    height=3.5cm
    ]
\addplot+[ybar] plot coordinates {(original, 0.164581) (model, 0.10844) (modularis, 0.1243938)};
\addplot+[ybar] plot coordinates {(original, 0.002756) (model, 0.01760949375) (modularis, 0.2900148)};
\addplot+[ybar] plot coordinates {(original, 2.692845) (model, 3.146234687) (modularis, 3.5681074)};
\addplot+[ybar] plot coordinates {(original, 0.4228548) (model, 0.6019256) (modularis, 0.9874098)};
\addplot+[ybar] plot coordinates {(original, 0.8226274) (model, 0.565304) (modularis, 0.6975944)};
\end{axis}
\end{tikzpicture}
\caption{Breakdown analysis for 4 (left) and 8 (right) machines}
\label{fig:micro}
\end{subfigure}
\begin{subfigure}[t]{0.33\textwidth}
\centering
\begin{tikzpicture}
\begin{axis}[
    xlabel={Number of machines},
    ylabel={Time[s]},
    xshift=2cm,
    ymin = 0,
    ymax = 18,
    xtick={0, 2, 3, 4, 5, 6, 7, 8},
    ylabel near ticks,
    legend pos=north east,
    legend entries={Monolithic, Modular},
    legend style={font=\tiny},
    width=4.5cm,
    height=4cm
]
\addplot coordinates {(2, 14.0600848) (3, 10.106813) (4, 7.7163566) (5, 6.3969024) (6, 5.5571646) (7,
4.93462) (8, 4.1056854)};
\addplot coordinates {(2, 16.6600974) (3, 11.5636978) (4, 9.3126192) (5, 8.1079808) (6, 6.594258) (7,
5.816244) (8, 5.6713998)};
\end{axis}
\end{tikzpicture}
\caption{Comparison across machines}
\label{fig:across}
\end{subfigure}
\vspace{-1em}
\caption{Modularis distributed join execution time per phase and compared to
    Monolithic design}
\label{fig:original}
\vspace{-1em}
\end{figure*}
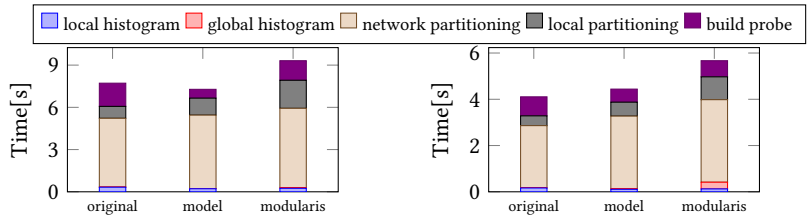
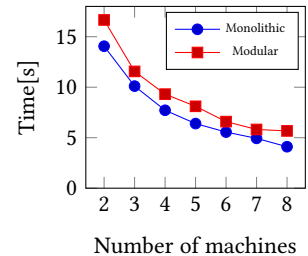
\label{sec:experiments}

\subsection{State-of-the-art distributed join}
\subsubsection{Implementation effort comparison}
Before we delve into a performance comparison of the execution of the
distributed hash join algorithm between Modularis and the original codebase, we
measure the implementation effort between the two approaches by using the
number of lines of code of each of them.
Although this metric is not always reliable, most of the
time it gives a good indication of the implementation effort. The operators that are used in the plan
according to Table~\ref{tab:legend} sum up to 1152 lines of code while the
original implementation adds to 1754 lines of code, leading to a 35\% reduction.
One can argue that this can be attributed to coding style but the main
take-away from this comparison is not only the size reduction but the extensibility
of our sub-operators. While to support other join types (e.g.\ semi-joins,
anti-joins) we only need to modify the \texttt{HashProbe} operator that consists
of 103 lines, the original codebase has to be replicated for every join
variant. Furthermore, the only operators that are platform-specific
used are: \texttt{MpiExecutor}, \texttt{MpiHistogram},
and \texttt{MpiExchange}, which sum up to 461 lines of code.
In contrast, the original monolithic code would have to be
rewritten from scratch if we wanted to change the target platform,
involving \SI{3.8}{\times} more code.

\subsubsection{Performance comparison}
We compare the distributed hash join code by Barthels et al.~\cite{Barthels2015},
which consists of
a monolithic operator, against our equivalent Modularis plan. For a fair
comparison, we extend the original code base with a similar materialization
operation to our \texttt{MaterializeRowVector} operator. The workload consists
of two relations with 2048 million tuples each.
Unless otherwise mentioned, we use a 1-to-1 correspondence between the keys in the inner and outer
relation.
This setup is consistent with the
workload used in the original paper \cite{Barthels2015} in the scale-out experiment (shown in Figure 7(a) in \cite{Barthels2015}).

We present our results in Figure~\ref{fig:original}. To understand how our
system performs, we also microbenchmark our sub-operators. These
microbenchmarks show the model performance that Modularis' \mbox{components} can achieve.
We compare the total runtime of these microbenchmarks (referred to as \emph{model})
against the entire Modularis query plan and the original code base.
We present our results for two machine configurations in Figure~\ref{fig:micro}.
We start our analysis by comparing the three execution times phase-by-phase.

Starting with the \emph{local histogram phase}, we observe that, compared to the original
code, both the model and the whole query plan have a small speedup.
We attribute this speedup to the fact that, as we mention in
Section~\ref{sec:complex-plans:partitioned-hash-join}, the local histogram
calculation is isolated in a small pipeline because its input has to be consumed
by multiple readers. This allows for compiler optimizations (e.g. automatic
usage of SIMD instructions, function inlining) that remove our sub-operators
abstractions. These optimizations
are not possible in larger pipelines, because in larger pipelines the compiler
cannot inline all the \texttt{next()} functions effectively.

The \emph{global histogram phase} has almost the same execution time
in our model and the original
code. However, in the full query plan and especially when the join is
executed on more machines, the total time is significantly larger. This is
associated with the \texttt{MPI\_Allreduce} function that calculates the
histogram, which is a collective operation that requires data from all the
processes. In case a process is stalled in a previous phase of the
algorithm, then every other process must wait until it has the
required data from it. In the original algorithm, this phenomenon is not present
because the histograms are calculated sequentially for both
relations. This also holds for the model. On the other hand, during
the execution of the join in Modularis, the global histogram calculation
happens in two distinct phases, one for each upstream path and a network
partitioning phase is between the two global histogram phases. Because the
network partitioning phase has a slight variation in its execution
time, it causes tail latencies for some processes during the calculation of
the global histogram of the second relation.

The \emph{network partitioning phase} is slower in the model and the
query plan than the original code base for two reasons.
First,
this operator is part of a large pipeline in our generated code and therefore, as
mentioned before, the compiler cannot perform all the possible optimizations and
remove all of our abstractions. To validate
this assumption and find the cause of the slowdown, we run the following
benchmark: We generate 1 billion integers and record the time that
\texttt{RowScan} needs to read them and compute their sum, compared to a simple
C++ program that does the same. \texttt{RowScan} needs about 1 second, whereas
the C++ program needs around 0.8 seconds.
The second reason for the slowdown is due to tail latencies because, as before,
the window allocation/synchronization function calls are collective operations.
In the original code base, they are called almost at the same time for both
relations but, in Modularis, they happen at different times,
one for each upstream path.

The \emph{local partitioning phase} is faster in the original code base than in the
model and the query plan. Part of the slowdown can be
again explained due to the complex pipelines that this phase belongs to, which
causes a comparative slowdown in the \texttt{RowScan} operator. This effect is
more eminent in the query plan because there the pipeline is even bigger than the
one in our microbenchmarks. Another cause of the slowdown is that in the
query plan, we cannot exactly isolate the local partitioning phase,
but we instead measure the whole sub-plan present in the first
\texttt{NestedMap} of Figure~\ref{fig:modular_join} and subsequently
subtract the one present in the second \texttt{NestedMap}.
This nested plan includes an extra materialization of the output partitions,
and the processing of metadata necessary for later phases of the algorithm.
Although the latter is not significantly compute-heavy, it attributes to a
small part of the slowdown present.

The \emph{build probe phase} is faster in the model, compared to the
original code base and the query plan. The slowdown in the first case
is explained by the fact that the \texttt{MaterializeRowVector} uses the
\texttt{realloc} function to request more memory, compared to the allocator
interface of the original code base. In the second case, build probe is
again part of a large pipeline. Therefore we lack some compiler
optimizations. On 8 machines, each process materializes fewer tuples
and requests less memory, and these effects are ameliorated.

Finally, in Figure~\ref{fig:across}, we depict the total runtime of the
monolithic operator compared to the Modularis plan for the distributed
join algorithm. Our modular integration is from 12 to 28\% slower, depending on
the number of machines used, due to the reasons explained before. However, the
operators present in this plan exhibit the advantage that they can be reused in
a variety of queries.

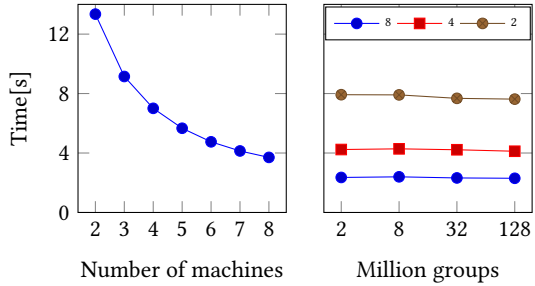
\begin{figure}[t]
\centering
\vspace{-2mm}
\begin{tikzpicture}
\begin{axis}[
    name=an1,
    xlabel={Number of machines},
    ylabel={Time[s]},
    xtick={2, 3, 4, 5, 6, 7, 8},
    ymin = 0,
    ymax = 14,
    ytick={0, 4, 8, 12},
    ylabel near ticks,
    width=4.35cm,
    height=4.35cm
]
 \addplot+ coordinates {(2, 13.3453132) (3, 9.1486338) (4, 7.0062696) (5, 5.666548) (6, 4.7525408) (7, 4.1385154) (8, 3.7051824)};
\end{axis}
\begin{axis}[
    at={(an1.south east)},
    xshift=0.5cm,
    xlabel={Million groups},
    legend entries={8, 4, 2},
    ymin = 0,
    ymax = 14,
    ytick={0, 4, 8, 12},
    yticklabels={},
    xtick={2, 8, 32, 128},
    legend style={at={(0.5, 0.99)},
    anchor=north,legend columns=-1, font=\tiny},
    xmode=log,
    log basis x={2},
    log ticks with fixed point,
    height=4.35cm,
    width=4.35cm
]

\addplot coordinates {(2, 2.3558316) (8, 2.3978786) (32, 2.323225) (128, 2.2999236)};
\addplot coordinates {(2, 4.2349282) (8, 4.28131922) (32, 4.2191104) (128, 4.119162)};
\addplot coordinates {(2, 7.9237124) (8, 7.911159) (32, 7.6785058) (128, 7.625359)};
\end{axis}

\end{tikzpicture}
\caption{Distributed \texttt{GROUP BY} runtime:
  varying cluster size with fixed key cardinality (left);
  varying key cardinality for different cluster sizes (right)}
\label{fig:groupby_singlekeys}
\end{figure}

\subsection{Distributed \texttt{\large GROUP BY}}
In this section, we run the distributed \texttt{GROUP BY} plan presented in
Section~\ref{sec:complex-plans:group-by}. We show our results in
Figure~\ref{fig:groupby_singlekeys}. On the left side of the Figure, we run the
plan across different machine configurations for a workload of 2048 unique
million keys. As expected, the total runtime decreases as the
algorithm load is distributed across more nodes. On the right side of the
Figure, we increase the number of distinct keys in the input (and hence the
number of groups in the result) and execute the
plan for three different machine configurations. Because the total execution
time is dominated by the network time and the materialization of the tuples,
the time is almost steady for each machine configuration. Overall,
we observe that Modularis performs grouping of billion of elements in a
few seconds without having to design a specialized monolithic operator for this
cause but mainly by reusing existing system components.

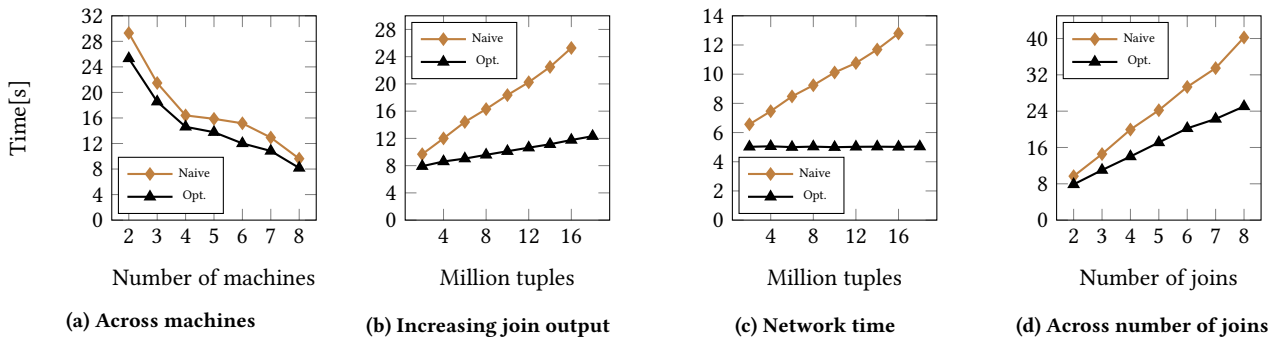
\begin{figure*}[b]
  \begin{subfigure}[t]{0.24\textwidth}
    \centering
    \begin{tikzpicture}[baseline={(xlabel.base)}]
\begin{axis}[
    name=ap1,
    xlabel={Number of machines},
    xlabel style={alias=xlabel},
    ylabel={Time[s]},
    ymin = 0,
    ymax = 32,
    xtick={2, 3, 4, 5, 6, 7, 8},
    ytick={0, 4, ..., 30, 32},
    legend pos=south west,
    legend entries={Naive, Opt.},
    legend style={font=\tiny},
    width=4.3cm,
    height=4.3cm,
]
\addplot[brown, mark=diamond*, thick] coordinates {(2, 29.2987248) (3, 21.4762894) (4, 16.401171) (5, 15.8618242) (6, 15.1642302) (7,
12.9320918) (8, 9.6229776)};
\addplot[black, mark=triangle*, thick] coordinates {(2, 25.3254454) (3, 18.5503132) (4, 14.6141892) (5, 13.7710004) (6, 12.0149374) (7,
10.8318528) (8, 8.1714366)};
\end{axis}
\end{tikzpicture}
\caption{Across machines}\label{fig:two_across}
\end{subfigure}
\begin{subfigure}[t]{0.24\textwidth}
\centering
\begin{tikzpicture}[baseline={(xlabel.base)}]
\begin{axis}[
    xlabel={Million tuples},
    xlabel style={alias=xlabel},
    ymin = 0,
    ymax = 30,
    xtick={4, 8, 12, 16, 20},
    ytick={0, 4, ..., 28},
    legend pos=north west,
    legend entries={Naive, Opt.},
    legend style={font=\tiny},
    width=4.3cm,
    height=4.3cm,
]
\addplot[brown, mark=diamond*, thick] coordinates {(2, 9.6791028) (4, 12.003459) (6, 14.4104596) (8, 16.3011804) (10, 18.3584482) (12, 20.2439144) (14, 22.4911942) (16, 25.278243)};
\addplot[black, mark=triangle*, thick] coordinates {(2, 7.919045) (4, 8.6094216) (6, 9.032907) (8, 9.5850214) (10, 10.1149386) (12,10.635747) (14, 11.1384616) (16, 11.7621006) (18, 12.3245494)};
\end{axis}
\end{tikzpicture}
\caption{Increasing join output}\label{fig:two_increasing}
\end{subfigure}
\begin{subfigure}[t]{0.24\textwidth}
\centering
\begin{tikzpicture}[baseline={(xlabel.base)}]
\begin{axis}[
    name=ap3,
    xlabel={Million tuples},
    xlabel style={alias=xlabel},
    xtick={4, 8, 12, 16, 20},
    ymin = 0,
    ymax = 14,
    ytick={0, 2, 4, ..., 12, 14},
    legend pos=south west,
    legend entries={Naive, Opt.},
    legend style={font=\tiny},
    width=4.3cm,
    height=4.3cm,
]
\addplot[brown, mark=diamond*, thick] coordinates {(2, 6.5639874) (4, 7.4628622) (6, 8.4762782) (8, 9.2414898) (10, 10.1155204) (12, 10.763866) (14, 11.6868908) (16, 12.7925022)};
\addplot[black, mark=triangle*, thick] coordinates {(2, 5.0228194) (4, 5.0581364) (6, 5.0016114) (8, 5.0355512) (10, 4.9966018) (12, 5.02294) (14, 5.0318638) (16, 5.0188762) (18, 5.0384534)};
\end{axis}
\end{tikzpicture}
\caption{Network time}\label{fig:network_increasing}
\end{subfigure}
\begin{subfigure}[t]{0.24\textwidth}
\centering
\begin{tikzpicture}[baseline={(xlabel.base)}]
\begin{axis}[
    name=ap4,
    xlabel={Number of joins},
    xlabel style={alias=xlabel},
    ymin = 0,
    ymax = 45,
    xtick={2, 3, ..., 9},
    ytick={0, 8, ..., 48},
    legend pos=north west,
    legend entries={Naive, Opt.},
    legend style={font=\tiny},
    width=4.3cm,
    height=4.3cm
]

\addplot[brown, mark=diamond*, thick]  coordinates {(2,9.6806818) (3, 14.5522736) (4, 19.9447402) (5, 24.2022282) (6, 29.3361678) (7, 33.47912) (8, 40.2318322)};
\addplot[black, mark=triangle*, thick] coordinates {(2, 7.8677998) (3, 11.0216796) (4, 14.010672) (5, 17.1065784) (6, 20.2144482) (7, 22.2856258) (8, 25.042437)};

\end{axis}
\end{tikzpicture}
    \caption{Across number of joins}\label{fig:sequences}
  \end{subfigure}
  \vspace{-1em}
  \caption{Sequences of joins in Modularis}
  \vspace{-1em}
\end{figure*}
\subsection{Sequences of joins}
Finally, we present the results of
executing sequences of joins. As a baseline, we use the naive version
of the plan that shuffles four relations through the network. We compare the
naive plan against an optimized one that shuffles only three. Both of these
versions are implemented in our system.
For this experiment, we use multiple relations with 2048 million tuples each,
similar to the relations used in \cite{Barthels2015}.
Additionally, we use a 1-to-1 correspondence between the keys in the inner and
outer relation unless otherwise mentioned.

Figure~\ref{fig:two_across} shows
the execution time when performing a sequence of two joins across several
machines with the two variants of the algorithm, \emph{naive}
and \emph{optimized}.
We observe that there is a constant speedup in the optimized version compared
to the baseline, which is partly due to the network shuffling of one less
relation and partly due to the materialization of only the final result instead
of materializing both the intermediate and the final join output. The execution
time has a sublinear speedup as the number of machines increases because the
tail latencies mentioned in the previous section in the network phases are even
more eminent. Because the network phases constitute a larger part of the
execution time of the optimized version, these tail latencies are the main
reason that the speedup between the baseline and the optimized version is
decreased as the number of machines increases.

However, the advantage of the optimized version is more conspicuous when the
first join has an increasing join output. We show the total runtime of such an
experiment across 8 machines in Figure~\ref{fig:two_increasing} and the time
spent on partitioning data through the network in
Figure~\ref{fig:network_increasing}. While the naive version
increases linearly with a high rate as the algorithm materializes and
shuffles through the network an extra relation that has an increasing size, the
optimized version has a sublinear increase in its total execution time. To
analyze further the cause of this time difference, we show in
Figure~\ref{fig:network_increasing} that for the optimized version the time
spent on shuffling data through the network is constant, as all three relations
are pre-partitioned at the beginning of the plan execution while the network
time is increasing linearly as more data are shuffled through the
network. In these two plots, we cannot execute the baseline
algorithm for more than 18 million tuples due to memory constraints.

Lastly, we compare the two versions while increasing the number of joins
performed. We show the results in Figure~\ref{fig:sequences}. The difference in
the total runtime between the two versions is proportional to the number of joins
because for $N$ joins, the optimized plan performs $N-1$ materializations
and $N + 1$ network shuffling phases less than the naive plan. This shows,
as in the \texttt{GROUP BY} case, that we can apply optimizations that have a
significant performance impact. This comes without rewriting the whole system from scratch,
as in the case of monolithic operators, but mainly by restructuring operators
across plans.

\section{Conclusion}
In this paper, we have proposed Modularis---an execution engine based on
sub-operators. After sketching the design principles that the sub-operators should
follow, we propose an initial set of sub-operators and demonstrate how they can be
combined to build traditional database operators such as a distributed hash join
or a \texttt{GROUP BY} query as well as more complex query plans like sequences
of joins and TPC-H queries. We show that by changing a small subset of our
sub-operators we can execute the same TPC-H query plans on \begin{revision}
diverse hardware platforms (RDMA, serverless clusters, smart storage).
\end{revision}
Through extensive experiments, we show that modularity reduces implementation
effort without requiring users to sacrifice performance.
Modularis is an order of magnitude faster than Presto, a data warehouse
supporting various storage layers and distributed setups, and more performant in
the majority of cases than SingleStore, an in-memory analytics SQL engine.
Modularis also outperforms Query-as-a-Service systems such as BigQuery and
Athena by simply changing the RDMA-specific operators with dedicated
serverless-based ones so that queries run on the cloud.

\bibliographystyle{ACM-Reference-Format}
\bibliography{biblio}


\begin{thebibliography}{71}


\ifx \showCODEN    \undefined \def \showCODEN     #1{\unskip}     \fi
\ifx \showDOI      \undefined \def \showDOI       #1{#1}\fi
\ifx \showISBNx    \undefined \def \showISBNx     #1{\unskip}     \fi
\ifx \showISBNxiii \undefined \def \showISBNxiii  #1{\unskip}     \fi
\ifx \showISSN     \undefined \def \showISSN      #1{\unskip}     \fi
\ifx \showLCCN     \undefined \def \showLCCN      #1{\unskip}     \fi
\ifx \shownote     \undefined \def \shownote      #1{#1}          \fi
\ifx \showarticletitle \undefined \def \showarticletitle #1{#1}   \fi
\ifx \showURL      \undefined \def \showURL       {\relax}        \fi
\providecommand\bibfield[2]{#2}
\providecommand\bibinfo[2]{#2}
\providecommand\natexlab[1]{#1}
\providecommand\showeprint[2][]{arXiv:#2}

\bibitem[\protect\citeauthoryear{??}{aqu}{2021}]%
        {aqua}
 \bibinfo{year}{Accessed 2021}\natexlab{}.
\newblock \bibinfo{title}{Amazon AQUA}.
\newblock
  \bibinfo{howpublished}{\url{https://aws.amazon.com/redshift/features/aqua/}}.
\newblock


\bibitem[\protect\citeauthoryear{??}{ath}{2021}]%
        {athena}
 \bibinfo{year}{Accessed 2021}\natexlab{}.
\newblock \bibinfo{title}{Athena external tables}.
\newblock
  \bibinfo{howpublished}{\url{https://docs.aws.amazon.com/athena/latest/ug/create-table.html}}.
\newblock


\bibitem[\protect\citeauthoryear{??}{big}{2021}]%
        {bigquery}
 \bibinfo{year}{Accessed 2021}\natexlab{}.
\newblock \bibinfo{title}{BigQuery external tables}.
\newblock
  \bibinfo{howpublished}{\url{https://cloud.google.com/bigquery/external-data-sources}}.
\newblock


\bibitem[\protect\citeauthoryear{??}{mut}{2021}]%
        {mutable}
 \bibinfo{year}{Accessed 2021}\natexlab{}.
\newblock \bibinfo{title}{Mutable project}.
\newblock
  \bibinfo{howpublished}{\url{https://bigdata.uni-saarland.de/projects/mutable/}}.
\newblock


\bibitem[\protect\citeauthoryear{??}{num}{2021}]%
        {numba}
 \bibinfo{year}{Accessed 2021}\natexlab{}.
\newblock \bibinfo{title}{Numba python package}.
\newblock \bibinfo{howpublished}{\url{http://numba.pydata.org/}}.
\newblock


\bibitem[\protect\citeauthoryear{??}{s3s}{2021}]%
        {s3select}
 \bibinfo{year}{Accessed 2021}\natexlab{}.
\newblock \bibinfo{title}{S3Select}.
\newblock
  \bibinfo{howpublished}{\url{https://aws.amazon.com/blogs/aws/s3-glacier-select/}}.
\newblock


\bibitem[\protect\citeauthoryear{??}{mem}{2021a}]%
        {memsqlcolumn}
 \bibinfo{year}{Accessed 2021}\natexlab{a}.
\newblock \bibinfo{title}{SingleStore column-store format}.
\newblock
  \bibinfo{howpublished}{\url{https://docs.memsql.com/v7.1/concepts/columnstore/}}.
\newblock


\bibitem[\protect\citeauthoryear{??}{mem}{2021b}]%
        {memsqlrow}
 \bibinfo{year}{Accessed 2021}\natexlab{b}.
\newblock \bibinfo{title}{SingleStore row-store format}.
\newblock
  \bibinfo{howpublished}{\url{https://docs.memsql.com/v7.1/concepts/rowstore/}}.
\newblock


\bibitem[\protect\citeauthoryear{Agrawal, Idicula, Raghavan, Vlachos,
  Govindaraju, tanathan Varadarajan, Balkesen, Giannikis, Roth, Agarwal, and
  Sedlar}{Agrawal et~al\mbox{.}}{2017}]%
        {AgrawalIRVGVBGR17}
\bibfield{author}{\bibinfo{person}{Sandeep~R. Agrawal}, \bibinfo{person}{Sam
  Idicula}, \bibinfo{person}{Arun Raghavan}, \bibinfo{person}{Evangelos
  Vlachos}, \bibinfo{person}{Venkatraman Govindaraju}, \bibinfo{person}{Venka\
  tanathan Varadarajan}, \bibinfo{person}{Cagri Balkesen},
  \bibinfo{person}{Georgios Giannikis}, \bibinfo{person}{Charlie Roth},
  \bibinfo{person}{Nipun Agarwal}, {and} \bibinfo{person}{Eric Sedlar}.}
  \bibinfo{year}{2017}\natexlab{}.
\newblock \showarticletitle{A Many-core Architecture for In-Memory Data
  Processing}. In \bibinfo{booktitle}{\emph{MICRO}}.
\newblock
\urldef\tempurl%
\url{https://doi.org/10.1145/3123939.3123985}
\showDOI{\tempurl}


\bibitem[\protect\citeauthoryear{Ao, Izhikevich, Voelker, and Porter}{Ao
  et~al\mbox{.}}{2018}]%
        {ao2018sprocket}
\bibfield{author}{\bibinfo{person}{Lixiang Ao}, \bibinfo{person}{Liz
  Izhikevich}, \bibinfo{person}{Geoffrey~M Voelker}, {and}
  \bibinfo{person}{George Porter}.} \bibinfo{year}{2018}\natexlab{}.
\newblock \showarticletitle{Sprocket: A serverless video processing framework}.
  In \bibinfo{booktitle}{\emph{SoCC}}. \bibinfo{pages}{263--274}.
\newblock
\urldef\tempurl%
\url{https://doi.org/10.1145/3267809.3267815}
\showDOI{\tempurl}


\bibitem[\protect\citeauthoryear{Balkesen, Teubner, Alonso, and
  {\"{O}}zsu}{Balkesen et~al\mbox{.}}{2013}]%
        {BalkesenTAO13}
\bibfield{author}{\bibinfo{person}{Cagri Balkesen}, \bibinfo{person}{Jens
  Teubner}, \bibinfo{person}{Gustavo Alonso}, {and} \bibinfo{person}{M.~Tamer
  {\"{O}}zsu}.} \bibinfo{year}{2013}\natexlab{}.
\newblock \showarticletitle{Main-Memory Hash Joins on Multi-Core CPUs: Tuning
  to the Underlying Hardware}. In \bibinfo{booktitle}{\emph{ICDE}}.
\newblock
\urldef\tempurl%
\url{https://doi.org/10.1109/ICDE.2013.6544839}
\showDOI{\tempurl}


\bibitem[\protect\citeauthoryear{Bandle and Giceva}{Bandle and Giceva}{2021}]%
        {bandle2021database}
\bibfield{author}{\bibinfo{person}{Maximilian Bandle} {and}
  \bibinfo{person}{Jana Giceva}.} \bibinfo{year}{2021}\natexlab{}.
\newblock \showarticletitle{Database Technology for the Masses: Sub-Operators
  as First-Class Entities}.
\newblock \bibinfo{journal}{\emph{PVLDB}} (\bibinfo{year}{2021}).
\newblock


\bibitem[\protect\citeauthoryear{Barthels, Loesing, Alonso, and
  Kossmann}{Barthels et~al\mbox{.}}{2015}]%
        {Barthels2015}
\bibfield{author}{\bibinfo{person}{Claude Barthels}, \bibinfo{person}{Simon
  Loesing}, \bibinfo{person}{Gustavo Alonso}, {and} \bibinfo{person}{Donald
  Kossmann}.} \bibinfo{year}{2015}\natexlab{}.
\newblock \showarticletitle{Rack-Scale In-Memory Join Processing Using RDMA}.
  In \bibinfo{booktitle}{\emph{SIGMOD}}.
\newblock
\urldef\tempurl%
\url{https://doi.org/10.1145/2723372.2750547}
\showDOI{\tempurl}


\bibitem[\protect\citeauthoryear{Barthels, M\"{u}ller, Schneider, Alonso, and
  Hoefler}{Barthels et~al\mbox{.}}{2017}]%
        {Barthels2017}
\bibfield{author}{\bibinfo{person}{Claude Barthels}, \bibinfo{person}{Ingo
  M\"{u}ller}, \bibinfo{person}{Timo Schneider}, \bibinfo{person}{Gustavo
  Alonso}, {and} \bibinfo{person}{Torsten Hoefler}.}
  \bibinfo{year}{2017}\natexlab{}.
\newblock \showarticletitle{Distributed Join Algorithms on Thousands of Cores}.
\newblock \bibinfo{journal}{\emph{PVLDB}} \bibinfo{volume}{10},
  \bibinfo{number}{5} (\bibinfo{year}{2017}).
\newblock
\urldef\tempurl%
\url{https://doi.org/10.14778/3055540.3055545}
\showDOI{\tempurl}


\bibitem[\protect\citeauthoryear{Barthels, M{\"{u}}ller, Taranov, Alonso, and
  Hoefler}{Barthels et~al\mbox{.}}{2019}]%
        {Barthels2019}
\bibfield{author}{\bibinfo{person}{Claude Barthels}, \bibinfo{person}{Ingo
  M{\"{u}}ller}, \bibinfo{person}{Konstantin Taranov}, \bibinfo{person}{Gustavo
  Alonso}, {and} \bibinfo{person}{Torsten Hoefler}.}
  \bibinfo{year}{2019}\natexlab{}.
\newblock \showarticletitle{{Strong consistency is not hard to get: Two-Phase
  Locking and Two-Phase Commit on Thousands of Cores}}.
\newblock \bibinfo{journal}{\emph{PVLDB}} \bibinfo{volume}{12},
  \bibinfo{number}{13} (\bibinfo{year}{2019}).
\newblock
\urldef\tempurl%
\url{https://doi.org/10.14778/3358701.3358702}
\showDOI{\tempurl}


\bibitem[\protect\citeauthoryear{Binnig, Crotty, Galakatos, Kraska, and
  Zamanian}{Binnig et~al\mbox{.}}{2016}]%
        {binnig2016end}
\bibfield{author}{\bibinfo{person}{Carsten Binnig}, \bibinfo{person}{Andrew
  Crotty}, \bibinfo{person}{Alex Galakatos}, \bibinfo{person}{Tim Kraska},
  {and} \bibinfo{person}{Erfan Zamanian}.} \bibinfo{year}{2016}\natexlab{}.
\newblock \showarticletitle{The end of slow networks: it's time for a
  redesign}.
\newblock \bibinfo{journal}{\emph{PVLDB}} \bibinfo{volume}{9},
  \bibinfo{number}{7} (\bibinfo{year}{2016}).
\newblock
\urldef\tempurl%
\url{https://doi.org/10.14778/2904483.2904485}
\showDOI{\tempurl}


\bibitem[\protect\citeauthoryear{Blanas, Koutris, and Sidi\-ropoulos}{Blanas
  et~al\mbox{.}}{2020}]%
        {Blanas2020}
\bibfield{author}{\bibinfo{person}{Spyros Blanas}, \bibinfo{person}{Paraschos
  Koutris}, {and} \bibinfo{person}{Anastasios Sidi\-ropoulos}.}
  \bibinfo{year}{2020}\natexlab{}.
\newblock \showarticletitle{{Topology-aware Parallel Data Processing: Models,
  Algorithms and Systems at Scale}}. In \bibinfo{booktitle}{\emph{CIDR}}.
\newblock


\bibitem[\protect\citeauthoryear{Boncz, Neumann, and Erling}{Boncz
  et~al\mbox{.}}{2013}]%
        {boncz2013tpc}
\bibfield{author}{\bibinfo{person}{Peter Boncz}, \bibinfo{person}{Thomas
  Neumann}, {and} \bibinfo{person}{Orri Erling}.}
  \bibinfo{year}{2013}\natexlab{}.
\newblock \showarticletitle{TPC-H analyzed: Hidden messages and lessons learned
  from an influential benchmark}. In \bibinfo{booktitle}{\emph{Technology
  Conference on Performance Evaluation and Benchmarking}}.
\newblock
\urldef\tempurl%
\url{https://doi.org/10.1007/978-3-319-04936-6_5}
\showDOI{\tempurl}


\bibitem[\protect\citeauthoryear{Carreira, Fonseca, Tumanov, Zhang, and
  Katz}{Carreira et~al\mbox{.}}{2019}]%
        {carreira2019cirrus}
\bibfield{author}{\bibinfo{person}{Joao Carreira}, \bibinfo{person}{Pedro
  Fonseca}, \bibinfo{person}{Alexey Tumanov}, \bibinfo{person}{Andrew Zhang},
  {and} \bibinfo{person}{Randy Katz}.} \bibinfo{year}{2019}\natexlab{}.
\newblock \showarticletitle{Cirrus: A serverless framework for end-to-end ml
  workflows}. In \bibinfo{booktitle}{\emph{SoCC}}. \bibinfo{pages}{13--24}.
\newblock
\urldef\tempurl%
\url{https://doi.org/10.1145/3357223.3362711}
\showDOI{\tempurl}


\bibitem[\protect\citeauthoryear{Cieslewicz and Ross}{Cieslewicz and
  Ross}{2007}]%
        {Cieslewicz2007}
\bibfield{author}{\bibinfo{person}{John Cieslewicz} {and} \bibinfo{person}{K.A.
  Ross}.} \bibinfo{year}{2007}\natexlab{}.
\newblock \showarticletitle{{Adaptive Aggregation on Chip Multiprocessors}}. In
  \bibinfo{booktitle}{\emph{VLDB}}.
\newblock


\bibitem[\protect\citeauthoryear{Crotty, Galakatos, Dursun, Kraska, Binnig,
  Cetintemel, and Zdonik}{Crotty et~al\mbox{.}}{2015}]%
        {Crotty2015tupleware}
\bibfield{author}{\bibinfo{person}{Andrew Crotty}, \bibinfo{person}{Alex
  Galakatos}, \bibinfo{person}{Kayhan Dursun}, \bibinfo{person}{Tim Kraska},
  \bibinfo{person}{Carsten Binnig}, \bibinfo{person}{Ugur Cetintemel}, {and}
  \bibinfo{person}{Stan Zdonik}.} \bibinfo{year}{2015}\natexlab{}.
\newblock \showarticletitle{An Architecture for Compiling UDF-centric
  Workflows}.
\newblock \bibinfo{journal}{\emph{PVLDB}} \bibinfo{volume}{8},
  \bibinfo{number}{12} (\bibinfo{year}{2015}).
\newblock
\urldef\tempurl%
\url{https://doi.org/10.14778/2824032.2824045}
\showDOI{\tempurl}


\bibitem[\protect\citeauthoryear{Dennl, Ziener, and Teich}{Dennl
  et~al\mbox{.}}{2012}]%
        {dennl2012fly}
\bibfield{author}{\bibinfo{person}{Christopher Dennl}, \bibinfo{person}{Daniel
  Ziener}, {and} \bibinfo{person}{Jurgen Teich}.}
  \bibinfo{year}{2012}\natexlab{}.
\newblock \showarticletitle{On-the-fly Composition of FPGA-Based SQL Query
  Accelerators Using a Partially Reconfigurable Module Library}. In
  \bibinfo{booktitle}{\emph{FCCM}}.
\newblock
\urldef\tempurl%
\url{https://doi.org/10.1109/FCCM.2012.18}
\showDOI{\tempurl}


\bibitem[\protect\citeauthoryear{Dewitt, Ghandeharizadeh, Schneider, Bricker,
  Hsiao, and Rasmussen}{Dewitt et~al\mbox{.}}{1990}]%
        {Dewitt1990gamma}
\bibfield{author}{\bibinfo{person}{D.~J. Dewitt}, \bibinfo{person}{S.
  Ghandeharizadeh}, \bibinfo{person}{D.~A. Schneider}, \bibinfo{person}{A.
  Bricker}, \bibinfo{person}{H.~I. Hsiao}, {and} \bibinfo{person}{R.
  Rasmussen}.} \bibinfo{year}{1990}\natexlab{}.
\newblock \showarticletitle{The Gamma Database Machine Project}.
\newblock \bibinfo{journal}{\emph{TKDE}} \bibinfo{volume}{2},
  \bibinfo{number}{1} (\bibinfo{year}{1990}).
\newblock
\urldef\tempurl%
\url{https://doi.org/10.1109/69.50905}
\showDOI{\tempurl}


\bibitem[\protect\citeauthoryear{Dittrich and Nix}{Dittrich and Nix}{2020}]%
        {dittrich2019case}
\bibfield{author}{\bibinfo{person}{Jens Dittrich} {and} \bibinfo{person}{Joris
  Nix}.} \bibinfo{year}{2020}\natexlab{}.
\newblock \showarticletitle{{The Case for Deep Query Optimisation}}. In
  \bibinfo{booktitle}{\emph{CIDR}}.
\newblock


\bibitem[\protect\citeauthoryear{Dittrich and Geppert}{Dittrich and
  Geppert}{2000}]%
        {dittrich2000component}
\bibfield{author}{\bibinfo{person}{Klaus~R Dittrich} {and}
  \bibinfo{person}{Andreas Geppert}.} \bibinfo{year}{2000}\natexlab{}.
\newblock \bibinfo{booktitle}{\emph{Component database systems}}.
\newblock \bibinfo{publisher}{Elsevier}.
\newblock


\bibitem[\protect\citeauthoryear{Dursun, Binnig, Cetintemel, Swart, and
  Gong}{Dursun et~al\mbox{.}}{2019}]%
        {dursun2019morsel}
\bibfield{author}{\bibinfo{person}{Kayhan Dursun}, \bibinfo{person}{Carsten
  Binnig}, \bibinfo{person}{Ugur Cetintemel}, \bibinfo{person}{Garret Swart},
  {and} \bibinfo{person}{Weiwei Gong}.} \bibinfo{year}{2019}\natexlab{}.
\newblock \showarticletitle{A Morsel-Driven Query Execution Engine for
  Heterogeneous Multi-Cores}.
\newblock \bibinfo{journal}{\emph{PVLDB}} \bibinfo{volume}{12},
  \bibinfo{number}{12} (\bibinfo{year}{2019}).
\newblock
\urldef\tempurl%
\url{https://doi.org/10.14778/3352063.3352137}
\showDOI{\tempurl}


\bibitem[\protect\citeauthoryear{Essertel, Tahboub, Wang, Decker, and
  Rompf}{Essertel et~al\mbox{.}}{2019}]%
        {flare_lantern2019}
\bibfield{author}{\bibinfo{person}{Gr{\'e}gory Essertel},
  \bibinfo{person}{Ruby~Y. Tahboub}, \bibinfo{person}{Fei Wang},
  \bibinfo{person}{James Decker}, {and} \bibinfo{person}{Tiark Rompf}.}
  \bibinfo{year}{2019}\natexlab{}.
\newblock \showarticletitle{Flare \& Lantern: Efficiently Swapping Horses
  Midstream}.
\newblock \bibinfo{journal}{\emph{PVLDB}} \bibinfo{volume}{12},
  \bibinfo{number}{12} (\bibinfo{year}{2019}).
\newblock
\urldef\tempurl%
\url{https://doi.org/10.14778/3352063.3352097}
\showDOI{\tempurl}


\bibitem[\protect\citeauthoryear{Essertel, Tahboub, Decker, Brown, Olukotun,
  and Rompf}{Essertel et~al\mbox{.}}{2018}]%
        {flare2018}
\bibfield{author}{\bibinfo{person}{Gr{\'e}gory~M. Essertel},
  \bibinfo{person}{Ruby~Y. Tahboub}, \bibinfo{person}{James~M. Decker},
  \bibinfo{person}{Kevin~J. Brown}, \bibinfo{person}{Kunle Olukotun}, {and}
  \bibinfo{person}{Tiark Rompf}.} \bibinfo{year}{2018}\natexlab{}.
\newblock \showarticletitle{Flare: Optimizing Apache Spark with Native
  Compilation for Scale-up Architectures and Medium-size Data}. In
  \bibinfo{booktitle}{\emph{OSDI}}.
\newblock


\bibitem[\protect\citeauthoryear{Fang, Zou, and Chien}{Fang
  et~al\mbox{.}}{2019}]%
        {fang2019accelerating}
\bibfield{author}{\bibinfo{person}{Yuanwei Fang}, \bibinfo{person}{Chen Zou},
  {and} \bibinfo{person}{Andrew~A Chien}.} \bibinfo{year}{2019}\natexlab{}.
\newblock \showarticletitle{Accelerating raw data analysis with the ACCORDA
  software and hardware architecture}.
\newblock \bibinfo{journal}{\emph{PVLDB}} \bibinfo{volume}{12},
  \bibinfo{number}{11} (\bibinfo{year}{2019}).
\newblock
\urldef\tempurl%
\url{https://doi.org/10.14778/3342263.3342634}
\showDOI{\tempurl}


\bibitem[\protect\citeauthoryear{Fouladi, Romero, Iter, Li, Chatterjee,
  Kozyrakis, Zaharia, and Winstein}{Fouladi et~al\mbox{.}}{2019}]%
        {fouladi2019laptop}
\bibfield{author}{\bibinfo{person}{Sadjad Fouladi}, \bibinfo{person}{Francisco
  Romero}, \bibinfo{person}{Dan Iter}, \bibinfo{person}{Qian Li},
  \bibinfo{person}{Shuvo Chatterjee}, \bibinfo{person}{Christos Kozyrakis},
  \bibinfo{person}{Matei Zaharia}, {and} \bibinfo{person}{Keith Winstein}.}
  \bibinfo{year}{2019}\natexlab{}.
\newblock \showarticletitle{From laptop to lambda: Outsourcing everyday jobs to
  thousands of transient functional containers}. In
  \bibinfo{booktitle}{\emph{USENIX ATC}}. \bibinfo{pages}{475--488}.
\newblock


\bibitem[\protect\citeauthoryear{Gerstenberger, Besta, and
  Hoefler}{Gerstenberger et~al\mbox{.}}{2018}]%
        {gerstenberger2014enabling}
\bibfield{author}{\bibinfo{person}{Robert Gerstenberger},
  \bibinfo{person}{Maciej Besta}, {and} \bibinfo{person}{Torsten Hoefler}.}
  \bibinfo{year}{2018}\natexlab{}.
\newblock \showarticletitle{Enabling Highly Scalable Remote Memory Access
  Programming with MPI-3 One Sided}.
\newblock \bibinfo{journal}{\emph{CACM}} \bibinfo{volume}{61},
  \bibinfo{number}{10} (\bibinfo{year}{2018}).
\newblock
\urldef\tempurl%
\url{https://doi.org/10.1145/3264413}
\showDOI{\tempurl}


\bibitem[\protect\citeauthoryear{Govindaraju, Gray, Kumar, and
  Manocha}{Govindaraju et~al\mbox{.}}{2006}]%
        {Govindaraju2006}
\bibfield{author}{\bibinfo{person}{Naga Govindaraju}, \bibinfo{person}{Jim
  Gray}, \bibinfo{person}{Ritesh Kumar}, {and} \bibinfo{person}{Dinesh
  Manocha}.} \bibinfo{year}{2006}\natexlab{}.
\newblock \showarticletitle{{GPUTeraSort: High Performance Graphics
  Co-processor Sorting for Large Database Management Naga}}. In
  \bibinfo{booktitle}{\emph{SIGMOD}}.
\newblock
\urldef\tempurl%
\url{https://doi.org/10.1145/1142473.1142511}
\showDOI{\tempurl}


\bibitem[\protect\citeauthoryear{Graefe}{Graefe}{1990}]%
        {Graefe1990}
\bibfield{author}{\bibinfo{person}{Goetz Graefe}.}
  \bibinfo{year}{1990}\natexlab{}.
\newblock \showarticletitle{{Encapsulation of Parallelism in the Volcano Query
  Processmg System}}. In \bibinfo{booktitle}{\emph{SIGMOD}}.
\newblock
\urldef\tempurl%
\url{https://doi.org/10.1145/93597.98720}
\showDOI{\tempurl}


\bibitem[\protect\citeauthoryear{He, Lu, Yang, Fang, Govindaraju, Luo, and
  Sander}{He et~al\mbox{.}}{2009}]%
        {He2009}
\bibfield{author}{\bibinfo{person}{Bingsheng He}, \bibinfo{person}{Mian Lu},
  \bibinfo{person}{Ke Yang}, \bibinfo{person}{Rui Fang},
  \bibinfo{person}{Naga~K. Govindaraju}, \bibinfo{person}{Qiong Luo}, {and}
  \bibinfo{person}{Pedro~V. Sander}.} \bibinfo{year}{2009}\natexlab{}.
\newblock \showarticletitle{{Relational Query Coprocessing on Graphics
  Processors}}.
\newblock \bibinfo{journal}{\emph{TDS}} \bibinfo{volume}{34},
  \bibinfo{number}{4} (\bibinfo{year}{2009}).
\newblock
\urldef\tempurl%
\url{https://doi.org/10.1145/1620585.1620588}
\showDOI{\tempurl}


\bibitem[\protect\citeauthoryear{He, Yang, Fang, Lu, Govindaraju, Luo, and
  Sander}{He et~al\mbox{.}}{2008}]%
        {He2008}
\bibfield{author}{\bibinfo{person}{Bingsheng He}, \bibinfo{person}{Ke Yang},
  \bibinfo{person}{Rui Fang}, \bibinfo{person}{Mian Lu},
  \bibinfo{person}{Naga~K. Govindaraju}, \bibinfo{person}{Qiong Luo}, {and}
  \bibinfo{person}{Pedro~V. Sander}.} \bibinfo{year}{2008}\natexlab{}.
\newblock \showarticletitle{{Relational Joins on Graphics Processors}}. In
  \bibinfo{booktitle}{\emph{SIGMOD}}.
\newblock
\urldef\tempurl%
\url{https://doi.org/10.1145/1376616.1376670}
\showDOI{\tempurl}


\bibitem[\protect\citeauthoryear{He, Sidler, Istv{\'{a}}n, and Alonso}{He
  et~al\mbox{.}}{2018}]%
        {HeSIA18}
\bibfield{author}{\bibinfo{person}{Zhenhao He}, \bibinfo{person}{David Sidler},
  \bibinfo{person}{Zsolt Istv{\'{a}}n}, {and} \bibinfo{person}{Gustavo
  Alonso}.} \bibinfo{year}{2018}\natexlab{}.
\newblock \showarticletitle{A Flexible K-Means Operator for Hybrid Databases}.
  In \bibinfo{booktitle}{\emph{FPL}}.
\newblock
\urldef\tempurl%
\url{https://doi.org/10.1109/FPL.2018.00069}
\showDOI{\tempurl}


\bibitem[\protect\citeauthoryear{Irmert, Daum, and Meyer-Wegener}{Irmert
  et~al\mbox{.}}{2008}]%
        {irmert2008new}
\bibfield{author}{\bibinfo{person}{Florian Irmert}, \bibinfo{person}{Michael
  Daum}, {and} \bibinfo{person}{Klaus Meyer-Wegener}.}
  \bibinfo{year}{2008}\natexlab{}.
\newblock \showarticletitle{A new approach to modular database systems}. In
  \bibinfo{booktitle}{\emph{Proceedings of the 2008 EDBT workshop on Software
  engineering for tailor-made data management}}. \bibinfo{pages}{40--44}.
\newblock


\bibitem[\protect\citeauthoryear{Jo, Bae, Yoon, Kang, Cho, Lee, and Jeong}{Jo
  et~al\mbox{.}}{2016}]%
        {jo2016yoursql}
\bibfield{author}{\bibinfo{person}{Insoon Jo}, \bibinfo{person}{Duck-Ho Bae},
  \bibinfo{person}{Andre~S Yoon}, \bibinfo{person}{Jeong-Uk Kang},
  \bibinfo{person}{Sangyeun Cho}, \bibinfo{person}{Daniel~DG Lee}, {and}
  \bibinfo{person}{Jaeheon Jeong}.} \bibinfo{year}{2016}\natexlab{}.
\newblock \showarticletitle{YourSQL: a high-performance database system
  leveraging in-storage computing}.
\newblock \bibinfo{journal}{\emph{PVLDB}} (\bibinfo{year}{2016}),
  \bibinfo{pages}{924--935}.
\newblock


\bibitem[\protect\citeauthoryear{Jonas, Pu, Venkataraman, Stoica, and
  Recht}{Jonas et~al\mbox{.}}{2017}]%
        {jonas2017occupy}
\bibfield{author}{\bibinfo{person}{Eric Jonas}, \bibinfo{person}{Qifan Pu},
  \bibinfo{person}{Shivaram Venkataraman}, \bibinfo{person}{Ion Stoica}, {and}
  \bibinfo{person}{Benjamin Recht}.} \bibinfo{year}{2017}\natexlab{}.
\newblock \showarticletitle{Occupy the cloud: Distributed computing for the
  99\%}. In \bibinfo{booktitle}{\emph{SoCC}}. \bibinfo{pages}{445--451}.
\newblock
\urldef\tempurl%
\url{https://doi.org/10.1145/3127479.3128601}
\showDOI{\tempurl}


\bibitem[\protect\citeauthoryear{Kara, Giceva, and Alonso}{Kara
  et~al\mbox{.}}{2017}]%
        {kaan17}
\bibfield{author}{\bibinfo{person}{Kaan Kara}, \bibinfo{person}{Jana Giceva},
  {and} \bibinfo{person}{Gustavo Alonso}.} \bibinfo{year}{2017}\natexlab{}.
\newblock \showarticletitle{FPGA-Based Data Partitioning}. In
  \bibinfo{booktitle}{\emph{SIGMOD}}.
\newblock
\urldef\tempurl%
\url{https://doi.org/10.1145/3035918.3035946}
\showDOI{\tempurl}


\bibitem[\protect\citeauthoryear{Kersten, Leis, Kemper, Neumann, Pavlo, and
  Boncz}{Kersten et~al\mbox{.}}{2018}]%
        {kersten2018everything}
\bibfield{author}{\bibinfo{person}{Timo Kersten}, \bibinfo{person}{Viktor
  Leis}, \bibinfo{person}{Alfons Kemper}, \bibinfo{person}{Thomas Neumann},
  \bibinfo{person}{Andrew Pavlo}, {and} \bibinfo{person}{Peter Boncz}.}
  \bibinfo{year}{2018}\natexlab{}.
\newblock \showarticletitle{Everything you always wanted to know about compiled
  and vectorized queries but were afraid to ask}.
\newblock \bibinfo{journal}{\emph{PVLDB}} \bibinfo{volume}{11},
  \bibinfo{number}{13} (\bibinfo{year}{2018}).
\newblock
\urldef\tempurl%
\url{https://doi.org/10.14778/3275366.3284966}
\showDOI{\tempurl}


\bibitem[\protect\citeauthoryear{Kim, Kaldewey, Lee, Sedlar, Nguyen, Satish,
  Chhugani, Di~Blas, and Dubey}{Kim et~al\mbox{.}}{2009}]%
        {kim2009sort}
\bibfield{author}{\bibinfo{person}{Changkyu Kim}, \bibinfo{person}{Tim
  Kaldewey}, \bibinfo{person}{Victor~W Lee}, \bibinfo{person}{Eric Sedlar},
  \bibinfo{person}{Anthony~D Nguyen}, \bibinfo{person}{Nadathur Satish},
  \bibinfo{person}{Jatin Chhugani}, \bibinfo{person}{Andrea Di~Blas}, {and}
  \bibinfo{person}{Pradeep Dubey}.} \bibinfo{year}{2009}\natexlab{}.
\newblock \showarticletitle{Sort vs. Hash Revisited: Fast Join Implementation
  on Modern Multi-Core CPUs}.
\newblock \bibinfo{journal}{\emph{PVLDB}} \bibinfo{volume}{2},
  \bibinfo{number}{2} (\bibinfo{year}{2009}).
\newblock
\urldef\tempurl%
\url{https://doi.org/10.14778/1687553.1687564}
\showDOI{\tempurl}


\bibitem[\protect\citeauthoryear{Kim and Lin}{Kim and Lin}{2018}]%
        {kim2018serverless}
\bibfield{author}{\bibinfo{person}{Youngbin Kim} {and} \bibinfo{person}{Jimmy
  Lin}.} \bibinfo{year}{2018}\natexlab{}.
\newblock \showarticletitle{Serverless data analytics with flint}. In
  \bibinfo{booktitle}{\emph{IEEE CLOUD}}. \bibinfo{pages}{451--455}.
\newblock
\urldef\tempurl%
\url{https://doi.org/10.1109/CLOUD.2018.00063}
\showDOI{\tempurl}


\bibitem[\protect\citeauthoryear{Klimovic, Wang, Kozyrakis, Stuedi, Pfefferle,
  and Trivedi}{Klimovic et~al\mbox{.}}{2018a}]%
        {klimovic2018understanding}
\bibfield{author}{\bibinfo{person}{Ana Klimovic}, \bibinfo{person}{Yawen Wang},
  \bibinfo{person}{Christos Kozyrakis}, \bibinfo{person}{Patrick Stuedi},
  \bibinfo{person}{Jonas Pfefferle}, {and} \bibinfo{person}{Animesh Trivedi}.}
  \bibinfo{year}{2018}\natexlab{a}.
\newblock \showarticletitle{Understanding ephemeral storage for serverless
  analytics}. In \bibinfo{booktitle}{\emph{ATC}}. \bibinfo{pages}{789--794}.
\newblock


\bibitem[\protect\citeauthoryear{Klimovic, Wang, Stuedi, Trivedi, Pfefferle,
  and Kozyrakis}{Klimovic et~al\mbox{.}}{2018b}]%
        {klimovic2018pocket}
\bibfield{author}{\bibinfo{person}{Ana Klimovic}, \bibinfo{person}{Yawen Wang},
  \bibinfo{person}{Patrick Stuedi}, \bibinfo{person}{Animesh Trivedi},
  \bibinfo{person}{Jonas Pfefferle}, {and} \bibinfo{person}{Christos
  Kozyrakis}.} \bibinfo{year}{2018}\natexlab{b}.
\newblock \showarticletitle{Pocket: Elastic ephemeral storage for serverless
  analytics}. In \bibinfo{booktitle}{\emph{OSDI}}. \bibinfo{pages}{427--444}.
\newblock


\bibitem[\protect\citeauthoryear{Klonatos, Koch, Rompf, and Chafi}{Klonatos
  et~al\mbox{.}}{2014}]%
        {legobase2014}
\bibfield{author}{\bibinfo{person}{Yannis Klonatos}, \bibinfo{person}{Christoph
  Koch}, \bibinfo{person}{Tiark Rompf}, {and} \bibinfo{person}{Hassan Chafi}.}
  \bibinfo{year}{2014}\natexlab{}.
\newblock \showarticletitle{Building Efficient Query Engines in a High-level
  Language}.
\newblock \bibinfo{journal}{\emph{PVLDB}} \bibinfo{volume}{7},
  \bibinfo{number}{10} (\bibinfo{year}{2014}).
\newblock
\urldef\tempurl%
\url{https://doi.org/10.14778/2732951.2732959}
\showDOI{\tempurl}


\bibitem[\protect\citeauthoryear{Kohn, Leis, and Neumann}{Kohn
  et~al\mbox{.}}{2021}]%
        {kohn2021building}
\bibfield{author}{\bibinfo{person}{Andr{\'e} Kohn}, \bibinfo{person}{Viktor
  Leis}, {and} \bibinfo{person}{Thomas Neumann}.}
  \bibinfo{year}{2021}\natexlab{}.
\newblock \showarticletitle{Building Advanced SQL Analytics From Low-Level Plan
  Operators}. In \bibinfo{booktitle}{\emph{SIGMOD}}.
  \bibinfo{pages}{1001--1013}.
\newblock


\bibitem[\protect\citeauthoryear{Leeka and Rajan}{Leeka and Rajan}{2019}]%
        {leeka19}
\bibfield{author}{\bibinfo{person}{Jyoti Leeka} {and} \bibinfo{person}{Kaushik
  Rajan}.} \bibinfo{year}{2019}\natexlab{}.
\newblock \showarticletitle{Incorporating Super-Operators in Big-Data Query
  Optimizers}.
\newblock \bibinfo{journal}{\emph{PVLDB}} \bibinfo{volume}{13},
  \bibinfo{number}{3} (\bibinfo{year}{2019}).
\newblock
\urldef\tempurl%
\url{https://doi.org/10.14778/3368289.3368299}
\showDOI{\tempurl}


\bibitem[\protect\citeauthoryear{Leis, Boncz, Kemper, and Neumann}{Leis
  et~al\mbox{.}}{2014}]%
        {Leis14}
\bibfield{author}{\bibinfo{person}{Viktor Leis}, \bibinfo{person}{Peter Boncz},
  \bibinfo{person}{Alfons Kemper}, {and} \bibinfo{person}{Thomas Neumann}.}
  \bibinfo{year}{2014}\natexlab{}.
\newblock \showarticletitle{Morsel-driven Parallelism: A NUMA-aware Query
  Evaluation Framework for the Many-core Age}. In
  \bibinfo{booktitle}{\emph{SIGMOD}}.
\newblock
\urldef\tempurl%
\url{https://doi.org/10.1145/2588555.2610507}
\showDOI{\tempurl}


\bibitem[\protect\citeauthoryear{Li, Das, Syamala, and Narasayya}{Li
  et~al\mbox{.}}{2016}]%
        {li2016accelerating}
\bibfield{author}{\bibinfo{person}{Feng Li}, \bibinfo{person}{Sudipto Das},
  \bibinfo{person}{Manoj Syamala}, {and} \bibinfo{person}{Vivek Narasayya}.}
  \bibinfo{year}{2016}\natexlab{}.
\newblock \showarticletitle{Accelerating Relational Databases by Leveraging
  Remote Memory and RDMA}. In \bibinfo{booktitle}{\emph{SIGMOD}}.
\newblock
\urldef\tempurl%
\url{https://doi.org/10.1145/2882903.2882949}
\showDOI{\tempurl}


\bibitem[\protect\citeauthoryear{Liu, Yin, and Blanas}{Liu
  et~al\mbox{.}}{2017}]%
        {liu2017design}
\bibfield{author}{\bibinfo{person}{Feilong Liu}, \bibinfo{person}{Lingyan Yin},
  {and} \bibinfo{person}{Spyros Blanas}.} \bibinfo{year}{2017}\natexlab{}.
\newblock \showarticletitle{Design and Evaluation of an RDMA-aware Data
  Shuffling Operator for Parallel Database Systems}. In
  \bibinfo{booktitle}{\emph{EuroSys}}.
\newblock
\urldef\tempurl%
\url{https://doi.org/10.1145/3064176.3064202}
\showDOI{\tempurl}


\bibitem[\protect\citeauthoryear{M{\"u}ller, Marroqu{\'\i}n, and
  Alonso}{M{\"u}ller et~al\mbox{.}}{2020}]%
        {muller2020lambada}
\bibfield{author}{\bibinfo{person}{Ingo M{\"u}ller}, \bibinfo{person}{Renato
  Marroqu{\'\i}n}, {and} \bibinfo{person}{Gustavo Alonso}.}
  \bibinfo{year}{2020}\natexlab{}.
\newblock \showarticletitle{Lambada: Interactive Data Analytics on Cold Data
  Using Serverless Cloud Infrastructure}. In
  \bibinfo{booktitle}{\emph{SIGMOD}}.
\newblock
\urldef\tempurl%
\url{https://doi.org/10.1145/3318464.3389758}
\showDOI{\tempurl}


\bibitem[\protect\citeauthoryear{M{\"{u}}ller, Sanders, Lacurie, Lehner, and
  F{\"{a}}rber}{M{\"{u}}ller et~al\mbox{.}}{2015}]%
        {Muller2015}
\bibfield{author}{\bibinfo{person}{Ingo M{\"{u}}ller}, \bibinfo{person}{Peter
  Sanders}, \bibinfo{person}{Arnaud Lacurie}, \bibinfo{person}{Wolfgang
  Lehner}, {and} \bibinfo{person}{Franz F{\"{a}}rber}.}
  \bibinfo{year}{2015}\natexlab{}.
\newblock \showarticletitle{{Cache-Efficient Aggregation: Hashing Is Sorting}}.
  In \bibinfo{booktitle}{\emph{SIGMOD}}.
\newblock
\urldef\tempurl%
\url{https://doi.org/10.1145/2723372.2747644}
\showDOI{\tempurl}


\bibitem[\protect\citeauthoryear{Neumann}{Neumann}{2011}]%
        {Neumann2011hyper}
\bibfield{author}{\bibinfo{person}{Thomas Neumann}.}
  \bibinfo{year}{2011}\natexlab{}.
\newblock \showarticletitle{Efficiently Compiling Efficient Query Plans for
  Modern Hardware}.
\newblock \bibinfo{journal}{\emph{PVLDB}} \bibinfo{volume}{4},
  \bibinfo{number}{9} (\bibinfo{year}{2011}).
\newblock
\urldef\tempurl%
\url{https://doi.org/10.14778/2002938.2002940}
\showDOI{\tempurl}


\bibitem[\protect\citeauthoryear{Palkar, Thomas, Narayanan, Thaker, Palamuttam,
  Negi, Shanbhag, Schwarzkopf, Pirk, Amarasinghe, Madden, and Zaharia}{Palkar
  et~al\mbox{.}}{2018}]%
        {Palkar2018weld}
\bibfield{author}{\bibinfo{person}{Shoumik Palkar}, \bibinfo{person}{James
  Thomas}, \bibinfo{person}{Deepak Narayanan}, \bibinfo{person}{Pratiksha
  Thaker}, \bibinfo{person}{Rahul Palamuttam}, \bibinfo{person}{Parimajan
  Negi}, \bibinfo{person}{Anil Shanbhag}, \bibinfo{person}{Malte Schwarzkopf},
  \bibinfo{person}{Holger Pirk}, \bibinfo{person}{Saman Amarasinghe},
  \bibinfo{person}{Samuel Madden}, {and} \bibinfo{person}{Matei Zaharia}.}
  \bibinfo{year}{2018}\natexlab{}.
\newblock \showarticletitle{Evaluating End-to-end Optimization for Data
  Analytics Applications in Weld}.
\newblock \bibinfo{journal}{\emph{PVLDB}} \bibinfo{volume}{11},
  \bibinfo{number}{9} (\bibinfo{year}{2018}).
\newblock
\urldef\tempurl%
\url{https://doi.org/10.14778/3213880.3213890}
\showDOI{\tempurl}


\bibitem[\protect\citeauthoryear{Perron, Castro~Fernandez, DeWitt, and
  Madden}{Perron et~al\mbox{.}}{2020}]%
        {perron2020starling}
\bibfield{author}{\bibinfo{person}{Matthew Perron}, \bibinfo{person}{Raul
  Castro~Fernandez}, \bibinfo{person}{David DeWitt}, {and}
  \bibinfo{person}{Samuel Madden}.} \bibinfo{year}{2020}\natexlab{}.
\newblock \showarticletitle{Starling: A Scalable Query Engine on Cloud
  Functions}. In \bibinfo{booktitle}{\emph{SIGMOD}}. \bibinfo{pages}{131--141}.
\newblock
\urldef\tempurl%
\url{https://doi.org/10.1145/3318464.3380609}
\showDOI{\tempurl}


\bibitem[\protect\citeauthoryear{Polychroniou, Raghavan, and Ross}{Polychroniou
  et~al\mbox{.}}{2015}]%
        {Polychroniou2015}
\bibfield{author}{\bibinfo{person}{Orestis Polychroniou}, \bibinfo{person}{Arun
  Raghavan}, {and} \bibinfo{person}{Kenneth~A. Ross}.}
  \bibinfo{year}{2015}\natexlab{}.
\newblock \showarticletitle{{Rethinking SIMD Vectorization for In-Memory
  Databases}}. In \bibinfo{booktitle}{\emph{SIGMOD}}.
\newblock
\urldef\tempurl%
\url{https://doi.org/10.1145/2723372.2747645}
\showDOI{\tempurl}


\bibitem[\protect\citeauthoryear{Polychroniou and Ross}{Polychroniou and
  Ross}{2014}]%
        {Polychroniou2014}
\bibfield{author}{\bibinfo{person}{Orestis Polychroniou} {and}
  \bibinfo{person}{Kenneth~A. Ross}.} \bibinfo{year}{2014}\natexlab{}.
\newblock \showarticletitle{{A Comprehensive Study of Main-Memory Partitioning
  and its Application to Large-Scale Comparison- and Radix-Sort}}. In
  \bibinfo{booktitle}{\emph{SIGMOD}}.
\newblock
\urldef\tempurl%
\url{https://doi.org/10.1145/2588555.2610522}
\showDOI{\tempurl}


\bibitem[\protect\citeauthoryear{Pu, Venkataraman, and Stoica}{Pu
  et~al\mbox{.}}{2019}]%
        {pu2019shuffling}
\bibfield{author}{\bibinfo{person}{Qifan Pu}, \bibinfo{person}{Shivaram
  Venkataraman}, {and} \bibinfo{person}{Ion Stoica}.}
  \bibinfo{year}{2019}\natexlab{}.
\newblock \showarticletitle{Shuffling, fast and slow: Scalable analytics on
  serverless infrastructure}. In \bibinfo{booktitle}{\emph{NSDI}}.
  \bibinfo{pages}{193--206}.
\newblock


\bibitem[\protect\citeauthoryear{R{\"o}diger, M{\"u}hlbauer, Kemper, and
  Neumann}{R{\"o}diger et~al\mbox{.}}{2015}]%
        {rodiger2015high}
\bibfield{author}{\bibinfo{person}{Wolf R{\"o}diger}, \bibinfo{person}{Tobias
  M{\"u}hlbauer}, \bibinfo{person}{Alfons Kemper}, {and}
  \bibinfo{person}{Thomas Neumann}.} \bibinfo{year}{2015}\natexlab{}.
\newblock \showarticletitle{High-Speed Query Processing over High-Speed
  Networks}.
\newblock \bibinfo{journal}{\emph{PVLDB}} \bibinfo{volume}{9},
  \bibinfo{number}{4} (\bibinfo{year}{2015}).
\newblock
\urldef\tempurl%
\url{https://doi.org/10.14778/2856318.2856319}
\showDOI{\tempurl}


\bibitem[\protect\citeauthoryear{Salama, Binnig, Kraska, Scherp, and
  Ziegler}{Salama et~al\mbox{.}}{2017}]%
        {salama2017rethinking}
\bibfield{author}{\bibinfo{person}{Abdallah Salama}, \bibinfo{person}{Carsten
  Binnig}, \bibinfo{person}{Tim Kraska}, \bibinfo{person}{Ansgar Scherp}, {and}
  \bibinfo{person}{Tobias Ziegler}.} \bibinfo{year}{2017}\natexlab{}.
\newblock \showarticletitle{Rethinking Distributed Query Execution on
  High-Speed Networks.}
\newblock \bibinfo{journal}{\emph{IEEE Data Eng. Bull.}} \bibinfo{volume}{40},
  \bibinfo{number}{1} (\bibinfo{year}{2017}).
\newblock
\urldef\tempurl%
\url{https://doi.org/10.14778/2904483.2904485}
\showDOI{\tempurl}


\bibitem[\protect\citeauthoryear{Samp{\'e}, Vernik, S{\'a}nchez-Artigas, and
  Garc{\'\i}a-L{\'o}pez}{Samp{\'e} et~al\mbox{.}}{2018}]%
        {sampe2018serverless}
\bibfield{author}{\bibinfo{person}{Josep Samp{\'e}}, \bibinfo{person}{Gil
  Vernik}, \bibinfo{person}{Marc S{\'a}nchez-Artigas}, {and}
  \bibinfo{person}{Pedro Garc{\'\i}a-L{\'o}pez}.}
  \bibinfo{year}{2018}\natexlab{}.
\newblock \showarticletitle{Serverless data analytics in the IBM cloud}. In
  \bibinfo{booktitle}{\emph{Proceedings of the 19th International Middleware
  Conference Industry}}. \bibinfo{pages}{1--8}.
\newblock
\urldef\tempurl%
\url{https://doi.org/10.1145/3284028.3284029}
\showDOI{\tempurl}


\bibitem[\protect\citeauthoryear{Schuhknecht, Khanchandani, and
  Dittrich}{Schuhknecht et~al\mbox{.}}{2015}]%
        {Schuhknecht2015}
\bibfield{author}{\bibinfo{person}{Felix~Martin Schuhknecht},
  \bibinfo{person}{Pankaj Khanchandani}, {and} \bibinfo{person}{Jens
  Dittrich}.} \bibinfo{year}{2015}\natexlab{}.
\newblock \showarticletitle{{On the Surprising Difficulty of Simple Things: the
  Case of Radix Partitioning}}.
\newblock \bibinfo{journal}{\emph{PVLDB}} \bibinfo{volume}{8},
  \bibinfo{number}{9} (\bibinfo{year}{2015}).
\newblock
\urldef\tempurl%
\url{https://doi.org/10.14778/2777598.2777602}
\showDOI{\tempurl}


\bibitem[\protect\citeauthoryear{Shankar, Krauth, Pu, Jonas, Venkataraman,
  Stoica, Recht, and Ragan-Kelley}{Shankar et~al\mbox{.}}{2018}]%
        {shankar2018numpywren}
\bibfield{author}{\bibinfo{person}{Vaishaal Shankar}, \bibinfo{person}{Karl
  Krauth}, \bibinfo{person}{Qifan Pu}, \bibinfo{person}{Eric Jonas},
  \bibinfo{person}{Shivaram Venkataraman}, \bibinfo{person}{Ion Stoica},
  \bibinfo{person}{Benjamin Recht}, {and} \bibinfo{person}{Jonathan
  Ragan-Kelley}.} \bibinfo{year}{2018}\natexlab{}.
\newblock \showarticletitle{Numpywren: Serverless linear algebra}.
\newblock \bibinfo{journal}{\emph{arXiv preprint arXiv:1810.09679}}
  (\bibinfo{year}{2018}).
\newblock


\bibitem[\protect\citeauthoryear{Sidler, Istv{\'{a}}n, Owaida, and
  Alonso}{Sidler et~al\mbox{.}}{2017}]%
        {SidlerIOA17}
\bibfield{author}{\bibinfo{person}{David Sidler}, \bibinfo{person}{Zsolt
  Istv{\'{a}}n}, \bibinfo{person}{Muhsen Owaida}, {and}
  \bibinfo{person}{Gustavo Alonso}.} \bibinfo{year}{2017}\natexlab{}.
\newblock \showarticletitle{Accelerating Pattern Matching Queries in Hybrid
  {CPU-FPGA} Architectures}. In \bibinfo{booktitle}{\emph{SIGMOD}}.
\newblock
\urldef\tempurl%
\url{https://doi.org/10.1145/3035918.3035954}
\showDOI{\tempurl}


\bibitem[\protect\citeauthoryear{Teubner and Mueller}{Teubner and
  Mueller}{2011}]%
        {Teubner2011}
\bibfield{author}{\bibinfo{person}{Jens Teubner} {and} \bibinfo{person}{Rene
  Mueller}.} \bibinfo{year}{2011}\natexlab{}.
\newblock \showarticletitle{{How Soccer Players Would do Stream Joins}}. In
  \bibinfo{booktitle}{\emph{SIGMOD}}.
\newblock
\urldef\tempurl%
\url{https://doi.org/10.1145/1989323.1989389}
\showDOI{\tempurl}


\bibitem[\protect\citeauthoryear{Woods, Istv{\'a}n, and Alonso}{Woods
  et~al\mbox{.}}{2014}]%
        {woods2014ibex}
\bibfield{author}{\bibinfo{person}{Louis Woods}, \bibinfo{person}{Zsolt
  Istv{\'a}n}, {and} \bibinfo{person}{Gustavo Alonso}.}
  \bibinfo{year}{2014}\natexlab{}.
\newblock \showarticletitle{Ibex: An intelligent storage engine with support
  for advanced sql offloading}.
\newblock \bibinfo{journal}{\emph{PVLDB}} (\bibinfo{year}{2014}),
  \bibinfo{pages}{963--974}.
\newblock


\bibitem[\protect\citeauthoryear{Yu, Youill, Woicik, Ghanem, Serafini,
  Aboulnaga, and Stonebraker}{Yu et~al\mbox{.}}{2020}]%
        {yu2020pushdowndb}
\bibfield{author}{\bibinfo{person}{Xiangyao Yu}, \bibinfo{person}{Matt Youill},
  \bibinfo{person}{Matthew Woicik}, \bibinfo{person}{Abdurrahman Ghanem},
  \bibinfo{person}{Marco Serafini}, \bibinfo{person}{Ashraf Aboulnaga}, {and}
  \bibinfo{person}{Michael Stonebraker}.} \bibinfo{year}{2020}\natexlab{}.
\newblock \showarticletitle{PushdownDB: Accelerating a DBMS using S3
  computation}. In \bibinfo{booktitle}{\emph{2020 IEEE ICDE}}.
\newblock


\bibitem[\protect\citeauthoryear{Zaharia, Chowdhury, Franklin, Shenker, Stoica,
  et~al\mbox{.}}{Zaharia et~al\mbox{.}}{2010}]%
        {zaharia2010spark}
\bibfield{author}{\bibinfo{person}{Matei Zaharia}, \bibinfo{person}{Mosharaf
  Chowdhury}, \bibinfo{person}{Michael~J Franklin}, \bibinfo{person}{Scott
  Shenker}, \bibinfo{person}{Ion Stoica}, {et~al\mbox{.}}}
  \bibinfo{year}{2010}\natexlab{}.
\newblock \showarticletitle{Spark: Cluster computing with working sets.}
\newblock \bibinfo{journal}{\emph{HotCloud}} \bibinfo{volume}{10},
  \bibinfo{number}{10-10} (\bibinfo{year}{2010}), \bibinfo{pages}{95}.
\newblock


\bibitem[\protect\citeauthoryear{Zamanian, Yu, Stonebraker, and
  Kraska}{Zamanian et~al\mbox{.}}{2019}]%
        {zamanian2019rethinking}
\bibfield{author}{\bibinfo{person}{Erfan Zamanian}, \bibinfo{person}{Xiangyao
  Yu}, \bibinfo{person}{Michael Stonebraker}, {and} \bibinfo{person}{Tim
  Kraska}.} \bibinfo{year}{2019}\natexlab{}.
\newblock \showarticletitle{Rethinking Database High Availability with RDMA
  Networks}.
\newblock \bibinfo{journal}{\emph{PVLDB}} \bibinfo{volume}{12},
  \bibinfo{number}{11} (\bibinfo{year}{2019}).
\newblock
\urldef\tempurl%
\url{https://doi.org/10.14778/3342263.3342639}
\showDOI{\tempurl}


\bibitem[\protect\citeauthoryear{Ziegler, Tumkur~Vani, Binnig, Fonseca, and
  Kraska}{Ziegler et~al\mbox{.}}{2019}]%
        {ziegler2019designing}
\bibfield{author}{\bibinfo{person}{Tobias Ziegler}, \bibinfo{person}{Sumukha
  Tumkur~Vani}, \bibinfo{person}{Carsten Binnig}, \bibinfo{person}{Rodrigo
  Fonseca}, {and} \bibinfo{person}{Tim Kraska}.}
  \bibinfo{year}{2019}\natexlab{}.
\newblock \showarticletitle{Designing Distributed Tree-based Index Structures
  for Fast RDMA-capable Networks}. In \bibinfo{booktitle}{\emph{SIGMOD}}.
\newblock
\urldef\tempurl%
\url{https://doi.org/10.1145/3299869.3300081}
\showDOI{\tempurl}


\end{thebibliography}

\end{document}